\documentclass[a4paper]{amsart}

\usepackage[T1]{fontenc}

\RequirePackage{amsmath}
\RequirePackage{bm}
\RequirePackage{amssymb}
\RequirePackage{upref}
\RequirePackage{amsthm}
\RequirePackage{enumerate}
\RequirePackage{pb-diagram}
\RequirePackage{amsfonts}
\RequirePackage[mathscr]{eucal}
\RequirePackage{verbatim}
\RequirePackage{xr}
\RequirePackage{graphicx}
\usepackage{calc}
\usepackage{xspace}
\RequirePackage{color}
\RequirePackage{ifthen}
\RequirePackage{fancybox}
\RequirePackage{mathrsfs}

\newcommand{\cf}{cf.\@\xspace}
\newcommand{\resp}{resp.\@\xspace}

\newcommand{\al}{\alpha}
\newcommand{\bet}{\beta}
\newcommand{\ga}{\gamma}
\newcommand{\de}{\delta }
\newcommand{\e}{\epsilon}

\newcommand{\f}{\varphi}

\newcommand{\lam}{\lambda}
\newcommand{\m}{\mu}

\newcommand{\om}{\omega}

\newcommand{\s}{\sigma}

\newcommand{\Clk}{\varGamma}
\newcommand{\D}{\varDelta}

\newcommand{\Lam}{\varLambda}
\newcommand{\Om}{\varOmega}

\newcommand{\socc}{{\mc S_0}}

\newcommand{\const}{\tup{const}}

\newcommand{\msp[1]}[1]{\mspace{#1mu}}

\newcommand{\R}[1][n+1]{{\protect\mathbb R}^{#1}}

\newcommand{\Cc}{{\protect\mathbb C}}

\newcommand{\N}{{\protect\mathbb N}}

\newcommand{\eR}{\stackrel{\lower1ex \hbox{\rule{6.5pt}{0.5pt}}}{\msp[3]\R[]}}
\newcommand{\eN}{\stackrel{\lower1ex \hbox{\rule{6.5pt}{0.5pt}}}{\msp[1]\N}}
\newcommand{\eO}{\stackrel{\lower1ex \hbox{\rule{6pt}{0.5pt}}}{\msc O}}

\DeclareMathOperator{\id}{id}

\DeclareMathOperator{\tr}{tr}

\DeclareMathOperator{\card}{card}

\newcommand\im{\implies}
\newcommand\ra{\rightarrow}

\newcommand\hra{\hookrightarrow}

\newcommand\whc[1]{\underset{#1}\rightharpoondown}

\newcommand\pa{\partial}
\newcommand\pde[2]{\frac {\partial#1}{\partial#2}}
 
\newcommand{\un}{\infty}
\newcommand{\A}{\forall}

\newcommand{\set}[2]{\{\,#1\colon #2\,\}}
\newcommand{\uu}{\cup}

\newcommand{\uuu}{\bigcup}

\newcommand{\uud}{ \stackrel{\lower 1ex \hbox {.}}{\uu}}
\newcommand{\uuud}[1]{ \stackrel{\lower 1ex \hbox {.}}{\uuu_{#1}}}
\newcommand\su{\subset}

\newcommand{\sminus}[1][28]{\raise 0.#1ex\hbox{$\scriptstyle\setminus$}}

\newcommand{\wed}{\wedge}

\newcommand{\abs}[1]{\lvert#1\rvert}

\newcommand{\norm}[1]{\lVert#1\rVert}

\newcommand{\nnorm}[1]{| \mspace{-2mu} |\mspace{-2mu}|#1| \mspace{-2mu}
|\mspace{-2mu}|}
\newcommand{\spd}[2]{\protect\langle #1,#2\protect\rangle}
\newcommand{\spdd}[2]{\protect\langle\protect\langle #1,#2\protect\rangle\protect\rangle}

\newcommand{\tit}{\textit}

\newcommand{\tup}{\textup}

\newcommand{\mc}{\protect\mathcal}
\newcommand{\msc}{\protect\mathscr}

\newcommand{\tlam}{\tilde\lam}
\newcommand{\tmu}{\tilde\mu}

\providecommand{\bysame}{\makebox[3em]{\hrulefill}\thinspace}

\newcommand{\bt}{\begin{thm}}
\newcommand{\bl}{\begin{lem}}
\newcommand{\bc}{\begin{cor}}
\newcommand{\bd}{\begin{definition}}
\newcommand{\bpp}{\begin{prop}}
\newcommand{\br}{\begin{rem}}
\newcommand{\bn}{\begin{note}}
\newcommand{\be}{\begin{ex}}
\newcommand{\bes}{\begin{exs}}
\newcommand{\bb}{\begin{example}}
\newcommand{\bbs}{\begin{examples}}
\newcommand{\ba}{\begin{axiom}}
\newcommand{\bas}{\begin{assumption}}

\newcommand{\et}{\end{thm}}
\newcommand{\el}{\end{lem}}
\newcommand{\ec}{\end{cor}}
\newcommand{\ed}{\end{definition}}
\newcommand{\epp}{\end{prop}}
\newcommand{\er}{\end{rem}}
\newcommand{\en}{\end{note}}
\newcommand{\ee}{\end{ex}}
\newcommand{\ees}{\end{exs}}
\newcommand{\eb}{\end{example}}
\newcommand{\ebs}{\end{examples}}
\newcommand{\ea}{\end{axiom}}
\newcommand{\eas}{\end{assumption}}

\newcommand{\bp}{\begin{proof}}
\newcommand{\ep}{\end{proof}}
\newcommand{\eps}{\renewcommand{\qed}{}\end{proof}}

\newcommand{\bal}{\begin{align}}

\newcommand{\bi}[1][1.]{\begin{enumerate}[\upshape #1]}
\newcommand{\bia}[1][(1)]{\begin{enumerate}[\upshape #1]}
\newcommand{\bin}[1][1]{\begin{enumerate}[\upshape\bfseries #1]}
\newcommand{\bir}[1][(i)]{\begin{enumerate}[\upshape #1]}
\newcommand{\bic}[1][(i)]{\begin{enumerate}[\upshape\hspace{2\cma}#1]}
\newcommand{\bis}[2][1.]{\begin{enumerate}[\upshape\hspace{#2\parindent}#1]}
\newcommand{\ei}{\end{enumerate}}

\newcommand\ndots{\raise 0.47ex \hbox {,}\hskip0.06em\cdots %
     \raise 0.47ex \hbox {,}\hskip0.06em} 

\newcommand{\q}{\quad}
\newcommand{\qq}{\qquad}

\newcommand\nd{\noindent}

\newskip\Csmallskipamount                                                
\Csmallskipamount=\smallskipamount
\newskip\Cmedskipamount
\Cmedskipamount=\medskipamount
\newskip\Cbigskipamount
\Cbigskipamount=\bigskipamount

\newcommand\cvs{\vspace\Csmallskipamount}   
\newcommand\cvm{\vspace\Cmedskipamount}

\newskip\csa
\csa=\smallskipamount

\newskip\cma
\cma=\medskipamount

\newskip\cba
\cba=\bigskipamount

\newdimen\spt
\newcommand\citem{\cvs\advance\itemno by
1{(\romannumeral\the\itemno})\hskip3pt}
\newcommand{\bitem}{\cvm\nd\advance\itemno by
1{\bf\the\itemno}\hspace{\cma}}

\newcount\itemno
\newcommand{\las}[1]{\label{S:#1}}
\newcommand{\lass}[1]{\label{SS:#1}}
\newcommand{\lae}[1]{\label{E:#1}}
\newcommand{\lat}[1]{\label{T:#1}}
\newcommand{\lal}[1]{\label{L:#1}}
\newcommand{\lad}[1]{\label{D:#1}}
\newcommand{\lac}[1]{\label{C:#1}}

\newcommand{\lar}[1]{\label{R:#1}}

\newcommand{\rt}[1]{Theorem~\ref{T:#1}}
\newcommand{\rl}[1]{Lemma~\ref{L:#1}}
\newcommand{\rd}[1]{Definition~\ref{D:#1}}

\newcommand{\rc}[1]{Corollary~\ref{C:#1}}

\newcommand{\rr}[1]{Remark~\ref{R:#1}}

\newcommand{\re}[1]{\eqref{E:#1}}
\newcommand{\frc}[1]{Corollary~\ref{C:#1} on page~\tup{\pageref{C:#1}}}
\newcommand{\frt}[1]{Theorem~\ref{T:#1} on page~\tup{\pageref{T:#1}}}
\newcommand{\frl}[1]{Lemma~\ref{L:#1} on page~\tup{\pageref{L:#1}}}
\newcommand{\frr}[1]{Remark~\ref{R:#1} on page~\tup{\pageref{R:#1}}}

\newcommand{\fre}[1]{\eqref{E:#1} on page~\tup{\pageref{E:#1}}}

\newcommand{\frs}[1]{Section~\ref{S:#1} on page~\tup{\pageref{S:#1}}}

\newskip\thmskip
\thmskip=\parindent

\newskip\hsk
\newenvironment{hinw}{\labelsep=0pt\begin{list}{}{\labelsep=0pt\itemindent=0pt\labelwidth=0pt\leftmargin=\parindent\rightmargin=0pt\partopsep=\cba}%
\item\it\nopagebreak\nopagebreak}%
{\end{list}}

\newcommand\bh{\begin{hinw}}
\newcommand{\eh}{\end{hinw}}

\newtheoremstyle{normal}
  {\cba}
  {\cba}
  {}
  {\thmskip}
  {\bfseries}
  {.}
  {\hsk}
  {}

\newtheoremstyle{abschnitt}
  {\cba}
  {\cba}
  {}
  {\thmskip}
  {\bfseries}
  {.}
  {\hsk}
  {}

\newtheoremstyle{italic}
  {\cba}
  {\cba}
  {\itshape}
  {\thmskip}
  {\bfseries}
  {.}
  {\hsk}
  {}

\newtheoremstyle{aufgaben}
  {\cba}
  {\cba}
  {}
  {}
  {\normalsize\bfseries}
  {.}
  {\hsk}
  {}

\newtheoremstyle{break}
  {\cba}
  {\cba}
  {\itshape}
  {}
  {\bfseries}
  {.}
  {\newline}
  {}

\theoremstyle{italic}
\newtheorem{thm}[subsection]{Theorem}
\newtheorem{lem}[subsection]{Lemma}
\newtheorem{prop}[subsection]{Proposition}
\newtheorem{cor}[subsection]{Corollary}

\theoremstyle{normal}
\newtheorem{rem}[subsection]{Remark}
\newtheorem{definition}[subsection]{Definition}
\newtheorem{example}[subsection]{Example}
\newtheorem{examples}[subsection]{Examples}
\newtheorem{ex}[subsection]{Exercise}
\newtheorem{note}[subsection]{}
\newtheorem{axiom}[subsection]{Axiom}
\newtheorem{assumption}[subsection]{Assumption}

\theoremstyle{aufgaben}
\newtheorem{exs}[subsection]{Exercises}

\numberwithin{equation}{section}
\numberwithin{figure}{section}

\newenvironment{textequation}[1][0.8]
{\begin{equation}
\begin{aligned}
\begin{minipage}{#1\linewidth}}
{\end{minipage}
\end{aligned}
\end{equation}
\ignorespacesafterend}

\newcommand{\btext}{\begin{textequation}}
\newcommand{\etext}{\end{textequation}}

\def\hinweis{\@startsection{subsection}{2}%
 \z@{0.7\linespacing\@plus 0.5\linespacing}{0.7\linespacing}%
{\normalfont\itshape\indent}}

\newcounter{hours}\newcounter{minutes}
\newcommand{\printtime}{%
\setcounter{hours}{\time/60}%
\setcounter{minutes}{\time-\value{hours}*60}%
\ifthenelse{\value{minutes}<10}{\thehours :0\theminutes}{\thehours:\theminutes}}

\usepackage[german,english]{babel}
\usepackage{graphicx}
\RequirePackage{amsmath}
\RequirePackage{bm}
\RequirePackage{amssymb}
\RequirePackage{upref}
\RequirePackage{amsthm}
\RequirePackage{enumerate}
\RequirePackage{pb-diagram}
\RequirePackage{amsfonts}
\RequirePackage[mathscr]{eucal}
\RequirePackage{verbatim}
\RequirePackage{xr}
\RequirePackage{graphicx}
\RequirePackage{mathrsfs}
\usepackage{calc}
\usepackage{xspace}
\usepackage{dsfont}

\makeatletter
\RequirePackage{color}
\newcommand{\ann}[1]{\renewcommand{\@makefnmark}{\mbox{$^{\color{red}{\@thefnmark}}$}}%
\footnote {#1}}
\makeatother

\RequirePackage{upref}
\RequirePackage{amsthm}
\RequirePackage{enumerate}
\usepackage[mathscr]{eucal}

\usepackage{xr-hyper}

\usepackage{calc}

\newlength{\oddsidemarginlength}
\newlength{\topmarginlength}

\newcounter{numberoflines}
\newcounter{tempcc}
\oddsidemargin=\oddsidemarginlength
\evensidemargin=\oddsidemargin
\topmargin=\topmarginlength
\usepackage[colorlinks=true,linkcolor=blue,citecolor=blue,urlcolor=blue]{hyperref}  

\begin{document}

\flushbottom


\title[The quantization of gravity]{The quantization of gravity: The quantization of the full Einstein equations}

\author{Claus Gerhardt}
\address{Ruprecht-Karls-Universit\"at, Institut f\"ur Angewandte Mathematik,
Im Neuenheimer Feld 205, 69120 Heidelberg, Germany}
\email{\href{mailto:gerhardt@math.uni-heidelberg.de}{gerhardt@math.uni-heidelberg.de}}
\urladdr{\href{http://www.math.uni-heidelberg.de/studinfo/gerhardt/}{http://www.math.uni-heidelberg.de/studinfo/gerhardt/}}

%


\date{\today}
%


\begin{abstract}
We quantized the full Einstein equations in a globally hyperbolic spacetime $N=N^{n+1}$, $n\ge 3$, and found solutions of the resulting hyperbolic equation in a fiber bundle $E$ which can be expressed as a product of spatial eigenfunctions (eigendistributions) and temporal eigenfunctions. The spatial eigenfunctions form a basis in an appropriate Hilbert space while the temporal eigenfunctions are solutions to a second order ordinary differential equation in $\R[]_+$. In case $n\ge 17$ and provided the cosmological constant  $\Lam$ is negative the temporal eigenfunctions are eigenfunctions of a self-adjoint operator $\hat H_0$ such that the eigenvalues are countable and the eigenfunctions form an orthonormal basis of a Hilbert space.
\end{abstract}

\maketitle

\tableofcontents

\setcounter{section}{0}
\section{Introduction}
General relativity is a Lagrangian theory and the canonical quantization of a Lagrangian theory is performed with the help of the Legendre transformation which would transform the Lagrangian theory to a an equivalent Hamiltonian theory provided that the Lagrangian is \tit{regular}, i.e., the second derivatives of the Lagrangian with respect to the time derivatives of the variables, which form a bilinear form, should be invertible. The Einstein-Hilbert Lagrangian is not regular. However, in a groundbreaking paper Arnowit, Deser and Misner (ADM) \cite{adm:old} proved that with the help of a global time function $x^0$ the Einstein-Hilbert functional could be expressed in a form which allowed to define a Hamiltonian $H$ and two constraints, the Hamilton constraint and the diffeomorphism constraint. Employing the Hamiltonian one could define the Hamilton equations and combined with the two constraints the resulting constrained Hamiltonian system was equivalent to the Einstein equations. Bryce DeWitt  used this constrained Hamiltonian system to perform a first canonical quantization of the Einstein equations in \cite{dewitt:gravity}. The Hamiltonian $H$ would be transformed to an operator $\hat H$ which would act on functions $u$ depending on Riemannian metrics $g_{ij}$ and the Hamilton constraint, which could be expressed as an equation,
\begin{equation}\lae{1.1.1}
H=0,
\end{equation}
would be transformed to the equation 
\begin{equation}\lae{1.2.1}
\hat Hu=0.
\end{equation}
The last equation is now known as the Wheeler-DeWitt equation. It could at first only be solved in highly symmetric cases like in the quantization of  Friedman universes, \cf \cite{misner:qc,unruh,kiefer_2022,moniz:qc,cg:qfriedman} and also the monographs \cite{kiefer:book,thiemann:book} and the bibliography therein. 

In \cite{cg:qgravity} we quantized a general globally hyperbolic spacetime $N=N^{n+1}$, $n\ge 3$, where $n$ is the space dimension, by using the afore mentioned papers \cite{adm:old,dewitt:gravity}. In that paper we first eliminated the diffeomorphism constraint by proving that the Einstein equations, which are the Euler-Lagrange equations of the Einstein-Hilbert functional, are equivalent to the Euler-Lagrange equations which are obtained by only considering Lorentzian metrics which split, i.e., they are of the form
\begin{equation}\lae{1.3.1}
d\bar s^2=-w^2 (dx^0)^2+g_{ij}(x^0,x) dx^idx^j,
\end{equation}
where the function $w>0$ and the Riemannian metrics $g_{ij}$ are arbitrary, \cf \cite[Theorem 3.2, p. 8]{cg:qgravity}. Let $G_{\al\bet}$, $0\le\al,\bet\le n$, be the Einstein tensor and $\Lam$ a cosmological constant. If only metrics of the form \re{1.3.1} are considered then the resulting Einstein equations can be split in a tangential part
\begin{equation}
G_{ij}+\Lam g_{ij}=0
\end{equation}
and a normal part
\begin{equation}
G_{\al\bet}\nu^\al\nu^\bet-\Lam=0,
\end{equation}
where $\nu=(\nu^\al)$ is a normal vector field to the Cauchy hypersurfaces
\begin{equation}
\{x^0=t\},\qq t\in x^0(N).
\end{equation}
The mixed Einstein equations are trivially satisfied since
\begin{equation}
G_{0j}=g_{0j}=0.
\end{equation}
The tangential Einstein equations are equivalent to the Hamilton equations, which are defined by the Hamiltonian $H$, and the normal equation is equivalent to the Hamilton constraint which can be expressed by the equation \re{1.1.1}. 

We also introduced a firm mathematical setting by quantizing a globally hyperbolic spacetime $N$ and working after the quantization in a fiber bundle $E$ with base space $\socc$, where $\socc$ was a Cauchy hypersurface of the quantized spacetime $N$. The fibers consisted of the Riemannian metrics defined in $\socc$. The quantized Hamiltonian $\hat H$ was a hyperbolic differential operator of second order in $E$ acting only in the fibers. We solved the Wheeler-DeWitt equation \re{1.2.1} in $E$, where $u=u(t,x,g_{ij})$, for given initial values, \cf \cite[Theorem 5.4, p.~18]{cg:qgravity}. Note that the Wheeler-DeWitt equation represents a quantization of the Hamilton condition, or equivalently, of the normal Einstein equation. The tangential Einstein equations have been ignored.  

In our paper \cite{cg:qgravity2b} and in the monograph  \cite{cg:qgravity-book} we finally quantized the full Einstein equations by incorporating the Hamilton condition in the Hamilton equations and we quantized this evolution equation. There are two possibilities how the Hamilton condition can be incorporated in the Hamilton equations and both  modified Hamilton equations combined with the original Hamilton equations   are equivalent to the full Einstein equations, \cf \cite[Theorem 1.3.3, p. 13, \& equ. 1.6.22, p. 41]{cg:qgravity-book}. After quantization of the modified Hamilton equations, however, the resulting hyperbolic equations are different: one equation, let us call it the first equation to give it name, is a hyperbolic equation where the elliptic parts---two Laplacians with respect to certain metrics---act both in the fibers as well as in the base space of a fiber bundle. The second equation is only a hyperbolic equation in the base space, since the  Laplacian acting in the fiber  had been  eliminated by the modification. 

The first equation has the form
\begin{equation}\lae{1.1}
-\D u-(n-1)\f \tilde\D u-\frac{n-2}2\f (R-2\Lam) u=0,
\end{equation}
\cf \cite[equ. (4.51)]{cg:qgravity2b} or \cite[equ. (1.4.88)]{cg:qgravity-book}, where the embellished Laplacian $\tilde\D u$ is the Laplacian in the base space $\socc$ with respect to the metric $g_{ij}$ if the function 
\begin{equation}
u\in C^\un_c(E,\Cc)
\end{equation}
is evaluated at 
\begin{equation}
(x, g_{ij}(x))\in E,
\end{equation}
or equivalently, after choosing appropriate coordinates in the fibers,
\begin{equation}\lae{1.4}
\begin{aligned}
&\frac n{16(n-1)} t^{-m}\frac\pa{\pa t}\big (t^m \pde ut\big )-t^{-2}\D_M u\\
&\q+t^{2-\frac4n}\{-(n-1) \D_{\s}u-\frac{n-2}2R_\s u\}+\frac{n-2}2 t^2\Lam u=0,
\end{aligned}
\end{equation}
where
\begin{equation}
m=\frac{(n-1)(n+2)}2\qq\wed\qq n=\dim\socc.
\end{equation}
The index $\s$ indicates that the corresponding geometric quantities are defined with respect to the metric $\s_{ij}\in M$, where  $M$ is the Cauchy hypersurface,
\begin{equation}
M=\{t=1\}. 
\end{equation}
The term $R_\s$ denotes the scalar curvature of the metric $\s_{ij}$ and $\Lam$ is a cosmological constant. By choosing a suitable atlas in the base space $\socc$, \cf \frl{3.1}, each fiber $M(x)$ consists of the positive definite matrices $\s_{ij}(x)$ satisfying 
\begin{equation}
\det \s_{ij}(x)=1,
\end{equation}
and hence, it is isometric to the symmetric space
\begin{equation}\lae{1.7.1}
SL(n,\R[])/SO(n)\equiv G/K.
\end{equation}
\cf \cite[equ.(5.17), p. 1123]{dewitt:gravity} and \cite[p. 3]{jorgenson:book}.

In \cite{cg:qgravity2b} and \cite{cg:qgravity-book} we could  solve the hyperbolic equation \re{1.4} only abstractly.  But because of the results in our paper \cite{cg:qgravity3} we are now able to apply separation of variables to express the solutions $u$ of \re{1.4} as a product of spatial and temporal eigenfunctions, or better, eigendistributions. There are three types of spatial eigenfunctions: First, the eigenfunctions of $-\D_M$ for which we choose the elements of the Fourier kernel $e_{\lam,b_0}$ such that 
\begin{equation}\lae{1.8}
-\D_M e_{\lam,b_0}=(\abs\lam^2+\abs\rho^2)e_{\lam,b_0},
\end{equation}
see \frs{3} for details, and then the eigenfunctions of the operator  
\begin{equation}\lae{1.9.1}
-(n-1)\D_\s-\frac{n-2}2 R_\s.
\end{equation}
While the operator in \re{1.8} acts in the fibers, and hence, the variables are the metrics $\s_{ij}\in M$, the operator in \re{1.9.1} is an elliptic differential operator of second order in $\socc$ for a fixed $\s_{ij}$. Thus, we have to specify a Riemannian metric $\s_{ij}$ in $\socc$ which is considered to be important either for physical or mathematical reasons. When a globally hyperbolic spacetime is quantized then $\socc$ is a Cauchy hypersurface, usually a coordinate slice, and it will be equipped with a Riemannian metric $\chi_{ij}$. It can be arranged that an arbitrary Riemannian metric $\chi_{ij}$ will be an element of $M$. Thus, our choice will be provided by the initial Cauchy hypersurface. In \cite{cg:uqtheory3b} we incorporated the Standard Model into our model and hence, we chose $\socc=\R[3]$ and $\chi_{ij}=\de_{ij}$.

When we quantized black holes, Schwarzschild-AdS or Kerr-AdS black holes, the interior region of a black hole can be considered to be a globally hyperbolic spacetime and the slices $\{r=\const\}$ are Cauchy hypersurfaces with induced Riemannian metrics $\chi_{ij}(r)$ (note that here  $r$ is  a label not a variable). If the event horizon is characterized by $r=r_0$ we proved that the Riemannian metrics $\chi_{ij}(r)$ converge to a Riemannian metric $\chi_{ij}(r_0)$ in an appropriate coordinate system. Thus, we chose $\socc$ to be the event horizon and $\chi_{ij}=\chi_{ij}(r_0)$. Moreover, $\socc$ could be written as  a product
\begin{equation}
\socc=\R[]\times M_0,
\end{equation}
where $M_0$ was a compact Riemannian manifold and $\chi$ a product metric
\begin{equation}\lae{1.12.1}
\chi=\de\otimes \bar\s,
\end{equation}
where $\de$ is the standard  "metric" in $\R[]$ and $\bar\s$ a Riemannian metric on $M_0$. 

Following the lead from the black holes we shall also assume in case of the quantization of a general globally hyperbolic spacetime $N=N^{n+1}$, $n\ge 3$,  that $\socc$ is a product
\begin{equation}\lae{1.13.1}
\socc=\R[n_1]\times M_0,
\end{equation}
at least topologically, and that  $M_0$ is a compact manifold of dimension
\begin{equation}
\dim M_0=n-n_1.
\end{equation}
If $N$  should be a mathematical model of our universe then we would choose $n_1=3$ and $M_0$ should be a compact manifold, hidden from our observation, of  fairly large dimension. Indeed we shall see that $n\ge 17$ would be preferable if at the same time the cosmological constant $\Lam$ would be negative. Moreover, assuming that $N$ should be equipped with an Einstein metric we would choose $M_0$ to be a Calabi-Yau manifold if $\Lam=0$, while in case of $\Lam<0$ $M_0$ should be a K\"ahler-Einstein space, and if $\Lam>0$ then $M_0$ is supposed to be a round sphere with a given radius. The metric $\s$ which we would use in the definition of the operator \re{1.9.1} would then be                                         
\begin{equation}\lae{1.15.1}
\s=\chi=\de\otimes \bar\s,
\end{equation}
where $\de$ would be the Euclidean metric in $\R[n_1]$ and $\bar\s$ the Riemannian metric in $M_0$. The differential operator in \re{1.9.1} would then have the form
\begin{equation}\lae{1.16.2}
-(n-1)\D_\de-(n-1)\D_{\bar\s}-\frac{n-2}2 R_{\bar\s},
\end{equation}
which would have eigenfunctions of the form 
\begin{equation}
\zeta\f
\end{equation}
where $\zeta$ is an eigenfunction of the Euclidean Laplacian and $\f$ an eigenfunction of the remaining part of the operator. Hence, we would have three types of spatial eigenfunctions which are well-known---both mathematically and physically---and their product will play the part of the spatial eigenfunctions of the hyperbolic equation \re{1.4}. The solution $u$ of that equation will then be of the form
\begin{equation}
u=wv\zeta\f
\end{equation}
where
\begin{equation}
v=e_{\lam,b_0}\circ [g_0]
\end{equation}
is an eigenfunction of $-\D_M$ satisfying
\begin{equation}
-\D_M v=(\abs\lam^2+\abs\rho^2)v
\end{equation}
and
\begin{equation}
v(\chi(x))=1\qq\A\, x\in \socc,
\end{equation}
for details we refer to the arguments following \frr{3.2}. The function $w$ depends only on $t$ and it will solve a second order differential equation (ODE). The functions $u$ will be evaluted at $(t,x,\chi)$. More precisely, we proved:
\bt\lat{1.1}
Assume that $\socc$ is a direct product as in \re{1.13.1} endowed with the metric $\chi$ in \re{1.15.1}. Then, a solution $u=u(x,t,\s_{ij})$ of the hyperbolic equation \re{1.4} can be expressed as a product of spatial eigenfunctions $v=v(\s_{ij})$, $\zeta=\zeta(y)$, $\f_k=\f_k(x)$, $k\in\N$, and  temporal eigenfunctions $w=w(t)$; $u$ is evaluated at $\s_{ij}=\chi_{ij}$, where
\begin{equation}
u=wv\zeta\f_k.
\end{equation}
The temporal eigenfunction $w$ is a solution  of the ODE 
\begin{equation}\lae{3.42.2}
\begin{aligned}
&\frac n{16(n-1)} t^{-m}\frac\pa{\pa t}\big (t^m \pde wt\big )+t^{-2}(\abs\lam^2+\rho^2)w\\
&\qq + t^{2-\frac4n}\{(n-1)\abs\xi^2+\bar\mu_k\}w+\frac{n-2}2 t^2\Lam w=0
\end{aligned}
\end{equation}
in $0<t<\un$.
\et

In \frs{5} we look at the case  $n\ge 17$ and $\Lam<0$ and prove that the equation \re{3.42.2} can be considered to be an implicit eigenvalue problem where $\Lam$ plays the part of the eigenvalue provided
\begin{equation}\lae{4.10.1}
\frac{16(n-1)}n\abs\lam^2<238.
\end{equation}
To understand  the corresponding theorem, we need a few remarks and definitions. First we multiply equation \re{3.42.2} by
\begin{equation}
\frac{16(n-1)}n,
\end{equation}
then we use the abbreviations
\begin{equation}\lae{4.2.2}
\mu_0=\frac{16(n-1)}n(\abs \lam^2+\abs\rho^2),
\end{equation}
\begin{equation}\lae{4.3.2}
m_1=\frac{16(n-1)}n\{(n-1)\abs\xi^2+\bar\mu_k\}
\end{equation}
and
\begin{equation}\lae{4.4.2}
m_2=\frac{8(n-1)(n-2)}n 
\end{equation}
and define for $w\in C^\un_c(\R[]_+)$ 
\begin{equation}
\hat Bw=-t^{-m}\frac\pa{\pa t}\big (t^m \pde wt\big )-t^{-2}\mu_0w.
\end{equation}
\br
Note that $\mu_0>0$ which would in general deprive of success any attempt to solve a meaningful eigenvalue problem for this operator. But if \re{4.10.1} is satisfied and $n\ge 17$, then it is possible to prove the following theorem in \frs{5}.
\er
\bt\lat{5.10.1}
There  are countably many solutions $(\Lam_i,w_i)$ of the implicit eigenvalue problem
\begin{equation}\lae{1.30} 
\hat Bw_i-m_2\Lam_it^{2}w_i=m_1 t^{2-\frac{4}n} w_i
\end{equation}
with eigenfunctions $w_i\in \mc{\hat H}_2$ such that
\begin{equation}
\Lam_i<\Lam_{i+1}<0\qq\A\,i\in\N,
\end{equation}
\begin{equation}
\lim_i\Lam_i=0,
\end{equation}
and their multiplicities are one.  
The transformed eigenfunctions 
\begin{equation}
\tilde w_i(t)=w_i(\lam_i^{\frac{n}{4(n-1)}}t),
\end{equation}
where
\begin{equation}
\lam_i=(-\Lam_i)^{-\frac {n-1}{n}},
\end{equation}
form a basis of $\mc{\hat H}_2$ and also of $L^2(\R[]_+,m)$.
\et
The equation \re{1.30} is the identical to  equation \re{3.42.2} if $\Lam$ is replaced by $\Lam_i$. The vector spaces $\mc{\hat H}_2$ and $L^2(\R[]_+,m)$ are Hilbert spaces which are defined later. 

However, if we consider  $\Lam<0$ to be a fixed cosmological constant and not a parameter which can also play the role of an implicit eigenvalue,  we have to use a different approach.

First, let us express equation \re{3.42.2} in the equivalent form 
\begin{equation}\lae{5.88.2}
\begin{aligned}
\hat\f_0^{-1}\bigg\{&-\frac\pa{\pa t}\big (t^m \pde wt\big )-t^{m-2}\mu_0w
- t^{m+2} m_2 \Lam w\bigg\}\\
&\qq\qq- \frac{16(n-1)}n\{(n-1)\abs\xi^2+\bar\mu_k\}  w=0,
\end{aligned}
\end{equation}
where
\begin{equation}
\hat\f_0(t)=t^{m+2-\frac4n}
\end{equation}
and where we used the definitions \re{4.2.2} and \re{4.4.2}. The term 
\begin{equation}
(n-1)\abs\xi^2+\bar\mu_k
\end{equation}
is an eigenvalue of the operator in \re{1.16.2}. $\abs \xi^2$ with $\xi\in \R[n_1]$ is a continuous eigenvalue while the sequence $\bar\mu_k$, $k\in\N$, satisfies the relations 
\begin{equation}\lae{3.39.2}
\bar\mu_0<\bar\mu_1\le\bar\mu_2\le\cdots
\end{equation}
and
\begin{equation}\lae{3.40.2}
\lim_{k\ra\un}\bar\mu_k=\un.
\end{equation}
The corresponding eigenfunctions $\f_k$ are smooth and the eigenspaces finite dimensional. 

On the other hand, the operator 
\begin{equation}\lae{5.91.2}
\hat H_0w\equiv \hat\f_0^{-1}\bigg\{-\frac\pa{\pa t}\big (t^m \pde wt\big )-t^{m-2}\mu_0w
- t^{m+2} m_2 \Lam w\bigg\}
\end{equation}
is  self-adjoint in the Hilbert space $\mc{\hat H}=L^2(\R[]_+,d\hat\mu)$, \cf \fre{5.80}, with a complete system of eigenfunctions $w_i$, $i\in\N$, and corresponding eigenvalues
\begin{equation}\lae{5.92.2}
0<\lam_0<\lam_1<\lam_2<\cdots
\end{equation}
The eigenspaces are all one dimensional and the ground state $w_0$ does not change sign, \cf \frr{5.8}. Thus, in order to solve equation \re{5.88.2} we have to find for each pair $(w_i,\lam_i)$ eigenvalues $\bar\mu_k$ and $\xi\in\R[n_1]$ such that 
\begin{equation}\lae{1.42.2}
\frac{16(n-1)}n\{(n-1)\abs\xi^2+\bar\mu_k\}=\lam_i.
\end{equation}
This is indeed possible provided either  $\bar\mu_0\le 0$ or
\begin{equation}
\abs{\Lam}^\frac{n-1}n\ge\bar\lam_0^{-1} \frac{16(n-1)}n\bar \mu_0,
\end{equation}
\cf \frc{5.16}.
Using the eigenvalues on the left-hand side of \re{1.42.2} and the corresponding eigenfunctions of the operator \re{1.16.2} we then define a self-adjoint operator $H_1$ in a Hilbert space $\mc H$ having the same eigenvalues $\lam_i$ as $\hat H_0$ but with higher finite multiplicities. Relabelling these eigenvalues to include the multiplicities and denoting them by $\tilde\lam_i$ they satisfy 
\begin{equation}
0<\tilde\lam_0\le\tilde\lam_1\le \cdots
\end{equation}
and 
\begin{equation}
\lim_{i\ra\un}\tilde\lam_i=\un.
\end{equation}
In \frs{6} we shall prove that the operator $e^{-\bet \hat H_0}$, $\bet>0$, is of trace class from which we conclude that $e^{-\bet H_1}$ is also of trace class. We are then in a similar situation as in \cite[Chapter 6.5]{cg:qgravity-book}, where we proved:
\bl
For any $\bet>0$ the operator
\begin{equation}
e^{-\bet H_1}
\end{equation}
is of trace class in $\mc H$, i.e., 
\begin{equation}
\tr (e^{-\bet H_1})=\sum_{i=0}^\un e^{-\beta \tilde\lam_i}<\un.
\end{equation}
 Let
\begin{equation}
\msc F\equiv\msc F_+(\mc H)
\end{equation}
be the symmetric Fock space generated by $\mc H$ and let
\begin{equation}
H=d\Clk(H_1)
\end{equation}
be the canonical extension of $H_1$ to $\msc F$. Then
\begin{equation}
e^{-\bet H}
\end{equation}
is also of trace class in $\msc F$
\begin{equation}
\tr(e^{-\bet H})=\prod_{i=0}^\un(1-e^{-\bet \tlam_i})^{-1}<\un.
\end{equation}
\el

\br
In \cite[Chapter 6.5]{cg:qgravity-book} we also used these results to define the partition function $Z$ by
\begin{equation}
Z=\tr(e^{-\bet H})=\prod_{i=0}^\un (1-e^{-\bet\tlam_i})^{-1}
\end{equation}
and the density operator $\rho$ in $\msc F$ by
\begin{equation}
\rho=Z^{-1}e^{-\bet H}
\end{equation}
such that
\begin{equation}
\tr \rho=1.
\end{equation}

The von Neumann entropy $S$ is then defined by
\begin{equation}
\begin{aligned}
S&=-\tr(\rho\log \rho)\\
&=\log Z+\bet Z^{-1}\tr (He^{-\bet H})\\
&=\log Z-\bet\pde{\log Z}\bet\\
&\equiv \log Z +\bet E,
\end{aligned}
\end{equation}
where $E$ is the average energy
\begin{equation}
E=\tr (H\rho).
\end{equation}
$E$ can be expressed in the form
\begin{equation}
E=\sum_{i=0}^\un \frac{\tlam_i}{e^{\bet\tlam_i}-1}.
\end{equation}
Here, we also set the Boltzmann constant
\begin{equation}
k_B=1.
\end{equation}
The parameter $\bet$ is supposed to be the inverse of the absolute temperature $T$
\begin{equation}
\bet=T^{-1}.
\end{equation}
For a more detailed analysis and especially for the dependence on $\Lam$ we refer  to  \cite[Chapter 6.5]{cg:qgravity-book}. 
\er

\br
Let us also mention that we use Planck units in this paper, i.e.,
\begin{equation}
c=G=k_B=\hbar=1.
\end{equation}
Moreover, the signature of a Lorentzian metric has the form $(-,+,\cdots,+)$.
\er

\section{Quantizing the full Einstein equations} \las{2} 

Let $N=N^{n+1}$, $n\ge 3$, be a globally hyperbolic Lorentzian manifold with metric $\bar g_{\al\bet}$, $0\le \al, \bet\le n$. The Einstein equations are Euler-Lagrange equations of the Einstein-Hilbert functional
\begin{equation}
\int_N (\bar R-\Lam),
\end{equation}
where $\bar R$ is the scalar curvature, $\Lam$ a cosmological constant and where we omitted the integration density in the integral. In order to apply a Hamiltonian description of general relativity, one usually defines a time function $x^0$ and considers the foliation of $N$ given by the slices
\begin{equation}
M(t)=\{x^0=t\}.
\end{equation}
We may, without loss of generality, assume that the spacetime metric splits
\begin{equation}\lae{1.3}
d\bar s^2=-w^2(dx^0)^2+g_{ij}(x^0,x)dx^idx^j,
\end{equation}
\cf \cite[Theorem 3.2]{cg:qgravity}. Then, the Einstein equations also split into a tangential part
\begin{equation}
G_{ij}+\Lam g_{ij}=0
\end{equation}
and a normal part
\begin{equation}
G_{\al\bet}\nu^\al\nu^\bet-\Lam=0,
\end{equation}
where the naming refers to the given foliation. For the tangential Einstein equations one can define equivalent Hamilton equations due to the groundbreaking paper by Arnowitt, Deser and Misner \cite{adm:old}. The normal Einstein equations can be expressed by the so-called Hamilton condition
\begin{equation}\lae{1.6}
\mc H=0,
\end{equation}
where $\mc H$ is the Hamiltonian used in defining the Hamilton equations. In the canonical quantization of gravity the Hamiltonian is transformed  to a partial differential operator of hyperbolic type $\hat{\mc H}$ and the possible quantum solutions of gravity are supposed to satisfy the so-called Wheeler-DeWitt equation
\begin{equation}\lae{1.7}
\hat{\mc H}u=0
\end{equation}
in an appropriate setting, i.e., only the Hamilton condition \re{1.6} has been quantized, or equivalently, the normal Einstein equation, while the tangential Einstein equations have been ignored.  

In \cite{cg:qgravity} we solved the equation \re{1.7} in a fiber bundle $E$ with base space $\socc$,
\begin{equation}
\socc=\{x^0=0\}\equiv M(0),
\end{equation}
and fibers $F(x)$, $x\in\socc$,
\begin{equation}\lae{2.9}
F(x)\su T^{0,2}_x(\socc),
\end{equation}
the elements of which are the positive definite symmetric tensors of order two, the Riemannian metrics in $\socc$. The hyperbolic operator $\hat{\mc H}$ is then expressed in the form 
\begin{equation}\lae{1.10}
\hat{\mc H}=-\D-(R-2\Lam)\f,
\end{equation}
where $\D$ is the Laplacian of the DeWitt metric given in the fibers, $R$ the scalar curvature of the metrics $g_{ij}(x)\in F(x)$, and $\f$ is defined by
\begin{equation}\lae{1.11}
\f^2=\frac{\det g_{ij}}{\det\rho_{ij}},
\end{equation}
where $\rho_{ij}$ is a fixed metric in $\socc$ such that instead of densities we are considering functions. The Wheeler-DeWitt equation could be solved in $E$ but only as an abstract hyperbolic equation. The solutions could not be split in corresponding spatial and temporal eigenfunctions.

The underlying mathematical reason for the difficulty was the presence of the term $R$ in the quantized equation, which prevents the application of separation of variables, since the metrics $g_{ij}$ are the spatial variables. In a recent paper \cite{cg:qgravity3} we overcame this difficulty by quantizing the Hamilton equations instead of the Hamilton condition. 
 
 As a result we obtained the equation
 \begin{equation}\lae{1.12}
-\D u=0
\end{equation}
in $E$, where the Laplacian is the Laplacian in \re{1.10}. The lower order terms of $\hat{\mc H}$ 
\begin{equation}
(R-2\Lam)\f
\end{equation}
were eliminated during the quantization process. However, the equation \re{1.12} is only valid provided $n\not=4$, since the resulting equation actually looks like
\begin{equation}
-(\frac n2-2)\D u=0.
\end{equation}
This restriction seems to be acceptable, since $n$ is the dimension of the base space $\socc$ which, by general consent, is assumed to be $n=3$. The fibers add additional dimensions to the quantized problem, namely,
\begin{equation}
\dim F=\frac {n(n+1)}2\equiv m+1.
\end{equation}
The fiber metric, the DeWitt metric, which is responsible for the Laplacian in \re{1.12} can be expressed in the form
\begin{equation}\lae{1.16.1}
ds^2=-\frac{16(n-1)}n dt^2+\f G_{AB}d\xi^Ad\xi^B,
\end{equation}
where the coordinate system is
\begin{equation}\lae{1.17}
(\xi^a)= (\xi^0,\xi^A)\equiv (t,\xi^A).
\end{equation}
The $(\xi^A)$, $1\le A\le m$, are coordinates for the hypersurface
\begin{equation}\lae{1.18}
M\equiv M(x)=\{(g_{ij}):t^4=\det g_{ij}(x)=1,\A\, x\in\socc\}.
\end{equation}
We also assumed that $\socc=\R[n]$ and that  the metric $\rho_{ij}$ in \re{1.11} is the Euclidean metric $\de_{ij}$. It is well-known that $M$ is a symmetric space
\begin{equation}\lae{1.19.1}
M=SL(n,\R[])/SO(n)\equiv G/K.
\end{equation}
It is also easily verified that the induced metric of $M$ in $E$ is isometric to the Riemannian metric of the coset space $G/K$.

Now, we were in a position to use separation of variables, namely, we wrote a solution of \re{1.12} in the form 
\begin{equation}
u=w(t)v(\xi^A),
\end{equation}
where $v$ is a spatial eigenfunction of the induced Laplacian of $M$ 
\begin{equation}\lae{1.21}
-\D_Mv\equiv -\D v=(\abs\lam^2+\abs\rho^2)v
\end{equation}
and $w$ is a temporal eigenfunction satisfying the ODE
\begin{equation}\lae{1.22}
\Ddot w+m t^{-1}\dot w+\mu_0 t^{-2}w=0
\end{equation}
with
\begin{equation}
\mu_0=\frac{16(n-1)}n(\abs \lam^2+\abs\rho^2).
\end{equation}

The eigenfunctions of the Laplacian in $G/K$ are well-known and we chose the kernel of the Fourier transform in $G/K$ in order to define the eigenfunctions. This choice also allowed us to use Fourier quantization similar to the Euclidean case such that the eigenfunctions are transformed to Dirac measures and the Laplacian to a multiplication operator in Fourier space. 

In the present paper we want to quantize the full Einstein equations by using a previous result, \cf \cite[Theorem 3.2]{cg:qgravity2b} or  \cite[Theorem 1.3.4]{cg:qgravity-book}, where we proved  that the full Einstein equations are equivalent  to the Hamilton equations and a scalar evolution equation which we obtained by incorporating the Hamilton condition into the right-hand side of the second Hamilton equations
and we quantized this evolution equation in fiber bundle $E$ with base space $\socc$ and fibers 
\begin{equation}
F(x)\in T_x^{0,2}(\socc),\qq \A x\in \socc,
\end{equation}
\cf \re{2.9}.

The quantization of the scalar evolution equation then yielded the following hyperbolic equation in $E$
\begin{equation}\lae{2.25}
-\D u-(n-1)\f \tilde\D u-\frac{n-2}2\f (R-2\Lam) u=0,
\end{equation}
\cf \cite[equ. (4.51)]{cg:qgravity2b} or \cite[equ. (1.4.88)]{cg:qgravity-book}. where the embellished Laplacian $\tilde\D u$ is the Laplacian in the base space $\socc$ with respect to the metric $g_{ij}$ if the function 
\begin{equation}
u\in C^\un_c(E,\Cc)
\end{equation}
is evaluated at 
\begin{equation}
(x, g_{ij}(x))\in E.
\end{equation}

Let us recall that the time function $t$ in \re{1.17} is defined by
\begin{equation}
t^2=\f
\end{equation}
and that $t$ is independent of $x$, \cf \cite[Lemma 4.1, p. 726]{cg:qgravity2b}, and, furthermore, that the fiber elements $g_{ij}(x)$ can be expressed as
\begin{equation}\lae{2.29}
g_{ij}(x)=t^\frac4n \s_{ij}(x),
\end{equation}
where the metrics $\s_{ij}(x)$ are elements of the fibers of the  subbundle
\begin{equation}\lae{2.30}
E_1=\{t=1\}\su E
\end{equation}
with fibers
\begin{equation}
M(x)\su F(x) \qq\A\, x\in \socc
\end{equation}
consisting of metrics $\s_{ij}(x)$ satisfying
\begin{equation}
\det \s_{ij}(x)=\det \rho_{ij}(x)\qq \A\, x\in \socc.
\end{equation}

Now, combining \re{2.29}, the definition of the fiber metric \re{1.16.1} and the relation between the scalar curvatures of conformal metrics the hyperbolic equation \re{2.25} can be expressed in the form
\begin{equation}\lae{2.33}
\begin{aligned}
&\frac n{16(n-1)} t^{-m}\frac\pa{\pa t}\big (t^m \pde ut\big )-t^{-2}\D_M u\\
&\q+t^{2-\frac4n}\{-(n-1) \D_{\s}u-\frac{n-2}2R_\s u\}+\frac{n-2}2 t^2\Lam u=0,
\end{aligned}
\end{equation}
where the index $\s$ indicates that the corresponding geometric quantities are defined with respect to the metric $\s_{ij}$.

In the following sections we shall solve equation \re{2.33} by employing separation of variables to obtain corresponding spatial and temporal eigenfunctions  or eigendistributions.

\section{Spatial eigenfunctions}\las 3

 Let us first look for spatial eigenfunctions of the operators
 \begin{equation}\lae{3.1}
-\D _M
\end{equation}
and
\begin{equation}\lae{3.2}
-(n-1)\D_\s-\frac{n-2}2 R_\s.
\end{equation}
In case of the Laplacian in \re{3.1} we would want to use the fact that each Cauchy hypersurface $M(x)$ is isometric to the symmetric space
\begin{equation}\lae{3.3}
SL(n,\R[])/SO(n)\equiv G/K
\end{equation}
provided
\begin{equation}\lae{3.4}
\det \rho_{ij}(x)=1.
\end{equation}
In our former papers \cite{cg:qgravity3} and  \cite{cg:uqtheory3b} we had chosen $\socc=\R[n]$ and
\begin{equation}
\begin{aligned}
\rho_{ij}=\de_{ij},
\end{aligned}
\end{equation}
i.e., the condition \re{3.4} had been automatically satisfied by choosing Euclidean coordinates. However, for the quantization of black holes this choice will not be possible since $\socc$ will then be the event horizon equipped with a non-flat metric.

To overcome this difficulty we need the following lemma:
\bl\lal{3.1}
Let $\socc$ be a Riemannian manifold of dimension $n\ge 2$ and of class $C^{k,\al}$ for $0\le k\in\N$ and $0<\al<1$, wheree $C^{k,\al}$ are the usual H\"older spaces, and let $\rho_{ij}$ be a metric of class $C^{k,\al}$ in $\socc$, then there exists an atlas $\{(x_\bet, U_\bet)\}$ of $C^{k+1,\al}$ charts such that the metric $\rho_{ij}$ expressed in an arbitrary chart $(x_\bet, U_\bet)$ satisfies
\begin{equation}\lae{3.6}
\det \rho_{ij}(x)=1\qq \A\, x\in x_\bet(U_\bet)\su \R[n].
\end{equation}
\el
\bp
We first prove \re{3.6} locally. Let $\rho_{ij}$ be a local expression of $\rho$ in coordinates $x=(x^i)$ and let $\tilde x=\tilde x(x)$ be a coordinate transformation and $\tilde\rho_{kl}$ be the corresponding expression for the metric $\rho$, then
\begin{equation}
\tilde\rho_{kl}=\rho_{ij} \frac{\pa x^i}{\pa \tilde x^k}\frac{\pa x^j}{\pa \tilde x^l}
\end{equation}
and
\begin{equation}
\det \tilde\rho_{kl}=\det\rho_{ij} 
\bigg|\frac{\pa x}{\pa \tilde x}\bigg|^2,
\end{equation}
where
\begin{equation}
\bigg|\frac{\pa x}{\pa \tilde x}\bigg|=\det \frac{\pa x^i}{\pa \tilde x^k},
\end{equation}
the Jacobi determinant.   

Let the coordinates $x=(x^i)$ be defined in an open set $\Om\su \R[n]$ with boundary $\pa\Om\in C^{k+1,\al}$, then, due to a result of Dacorogna and Moser, there exists a diffeomorphism $y=y(x)$, $y\in C^{k+1,\al}(\bar\Om,\R[n])$ such that 
\begin{equation}
\begin{aligned}
\bigg|\frac{\pa y}{\pa x}\bigg|&=\lam \sqrt{\det \rho_{ij}}\qq \tup{in}\;\Om,\\
y(x)&=x\qq\qq\msp[30]\tup{in}\; \pa\Om,
\end{aligned}
\end{equation}
where
\begin{equation}
\lam=\frac{\int_\Om dx}{\int_\Om\sqrt{\det \rho_{ij}\,dx}},
\end{equation}
\cf \cite[Theorem 1' \& Remark, p. 4]{moser:jacobian}. 

Hence, the diffeomorphism
\begin{equation}
\tilde x=\lam^\frac1n y
\end{equation}
satisfies
\begin{equation}
\bigg|\frac{\pa \tilde x}{\pa x}\bigg|=\sqrt{\det \rho_{ij}}, 
\end{equation}
or equivalently,
\begin{equation}
\det \tilde\rho_{kl}=\det\rho_{ij} 
\bigg|\frac{\pa x}{\pa \tilde x}\bigg|^2=1,
\end{equation}
where $\tilde \rho_{kl}$ are the coordinate expressions of $\rho$ in the coordinates $\tilde x$.

From the local result we easily infer the existence of an atlas consisting of local charts with that property.
\ep

Thus, we are able to identify the fiber $M(x)$ with the symmetric space $G/K$ in \re{3.3} and we may choose the elements of the Fourier kernel $e_{\lam,b_0}$ as eigenfunctions of $-\D_M$ such that
 \begin{equation}\lae{3.15}
-\D_M e_{\lam,b_0}=(\abs\lam^2+\abs\rho^2)e_{\lam,b_0},
\end{equation}
see \cite[Chapter III]{helgason:ga} and \cite[Section 5]{cg:qgravity3} for details, where
\begin{equation}
\abs\rho^2=\frac1{12}(n-1)^2n,
\end{equation}
\cf \cite[equ. (5.40)]{cg:qgravity3}. Here, $\lam$ is an abbreviation for $\lam \al$, where $\al\in(\R[n-1])^*$ is a character representing an elementary graviton and $\lam\in\R[]_+$. There are
\begin{equation}
\al=\begin{cases}
\al_i,&1\le i\le n-1\\
\al_{ij},&1\le i<j\le n
\end{cases}
\end{equation}
special characters. These characters are normalized to have $\norm\al=1$. They correspond to the degrees of freedom in choosing the entries of a metric $g_{ij}$ satisfying
\begin{equation}
\det g_{ij}=1.
\end{equation}

\br\lar{3.2}
Due to the scalar curvature term $R_\s$ in equation \re{3.2} it is evident that spatial eigenfunctions for this operator cannot be defined on the full subbundle $E_1$, \cf \fre{2.30}, but only for a fixed metric $\s_{ij}\in M$, if $R_\s=\const$ maybe for that class of metrics. However, in general, we cannot assume that the scalar curvature is constant, since we shall have to pick a metric $\chi_{ij}$ that is a natural metric determined by the underlying spacetime which has been quantized. In case of a black hole $\chi_{ij}$ will be a metric on the event horizon. Now, let us recall that $\chi_{ij}$ should belong to fibers of the subbundle $E_1$, hence, we have to choose $\rho_{ij}$, which is still arbitrary but fixed, to be equal to $\chi_{ij}$ 
\begin{equation}\lae{3.19}
\rho_{ij}=\chi_{ij}.
\end{equation}
\er

Thus, we evaluate the spatial eigenfunctions at 
\begin{equation}
(x,\chi_{ij}(x))\qq\A\, x\in \socc,
\end{equation}
especially also $e_{\lam,b_0}$, i.e., 
\begin{equation}
e_{\lam,b_0}(\chi_{ij}(x))
\end{equation}
may not depend on $x$ explicitly.  Now, it is well known that
\begin{equation}
e_{\lam,b_0}(\de_{ij}(x))=1\qq \A\, x\in \socc
\end{equation}
and the Laplacian $\D_M$ is invariant under the action of $G$ on $M$. The action of $g\in M$ on $\s\in M$ is defined by
\begin{equation}
[g]\s=g\s g^*,
\end{equation}
where $g^*$ is the transposed matrix. Since every $\s\in M$ is also an element of $G$ we  conclude, by choosing
\begin{equation}\lae{3.24}
g=g_0\equiv \sqrt{\chi^{-1}},
\end{equation}
that
\begin{equation}
[g_0]\chi=\id=(\de_{ij}),
\end{equation}
and, furthermore, that the function
\begin{equation}
v=e_{\lam,b_0}\circ [g_0]
\end{equation}
is an eigenfunction of $-\D_M$ satisfying
\begin{equation}\lae{3.27}
-\D_M v=(\abs\lam^2+\abs\rho^2)v
\end{equation}
and
\begin{equation}\lae{3.28}
v(\chi(x))=1\qq\A\, x\in \socc.
\end{equation}
Let us summarize these results in
\bt\lat{3.3}
Let $e_{\lam, b_0}$ be an eigenfunction of $-\D_M$ as in \re{3.15} and let $g_0$ be defined as in \re{3.24}, then 
\begin{equation}
v=e_{\lam,b_0}\circ [g_0]
\end{equation}
is an eigenfunction of $-\D_M$ satisfying \re{3.27} as well as \re{3.28}. 
\et

Next, let us consider the operator in \re{3.2} with $\s=\chi$. We furthermore  assume that $\socc$ is a direct product,
\begin{equation}\lae{3.30}
\socc=\R[n_1]\times M_0,
\end{equation}
where $M_0$ is a smooth, compact and connected manifold of dimension $n-n_1$,
\begin{equation}
\dim M_0=n-n_1\equiv n_0.
\end{equation}
The metric $\chi_{ij}$ is then supposed to be a metric product,
\begin{equation}\lae{3.32}
\chi=\de\otimes\bar\s,
\end{equation}
where $\de$ is the Euclidean metric in $\R[n_1]$ and $\bar\s$ a Riemannian metric in $M_0$. In case of a black hole $n_1$ will be equal to $1$.  

Since the scalar curvature of the product metric $\chi$ is equal to the scalar curvature of $\bar\s$,
\begin{equation}
R_\chi=R_{\bar\s},
\end{equation}
the operator in \re{3.2} can be expressed in the form
\begin{equation}\lae{3.34}
-(n-1)\D_\de-(n-1)\D_{\bar\s}-\frac{n-2}2 R_{\bar\s}.
\end{equation}
Hence, the corresponding eigenfunctions can be written as a product
\begin{equation}
\zeta \f,
\end{equation}
where $\zeta$ is defined in $\R[n_1]$,
\begin{equation}
\zeta(y)=e^{i\spd \xi y}\qq  \xi, y\in\R[n_1],
\end{equation}
such that
\begin{equation}
-\D_\de\zeta=\abs\xi^2\zeta,
\end{equation}
while $\f\in C^\un(M_0)$ is an eigenfunction of the operator
\begin{equation}\lae{3.38}
A=-(n-1)\D_{\bar\s}-\frac{n-2}2 R_{\bar\s}.
\end{equation}
Since $M_0$ is compact it is well-known that $A$ is self-adjoint with countably many eigenvalues $\bar\mu_k$, $k\in \N$, which are ordered
\begin{equation}\lae{3.39}
\bar\mu_0<\bar\mu_1\le\bar\mu_2\le\cdots
\end{equation}
satisfying
\begin{equation}\lae{3.40}
\lim_{k\ra\un}\bar\mu_k=\un.
\end{equation}
The corresponding eigenfunctions $\f_k$ are smooth and the eigenspaces finite dimensional. The eigenspace belonging to $\bar\mu_0$ is one dimensional and $\f_0$ never vanishes, i.e., if we consider $\f_0$ to be real valued it will either be strictly positive or negative.

Let us summarize the results we proved so far in the following theorem:
\bt\lat{3.4}
Assume that $\socc$ is a direct product as in \re{3.30} endowed with the metric $\chi$ in \re{3.32}. Then, a solution $u=u(x,t,\s_{ij})$ of the hyperbolic equation \fre{2.33} can be expressed as a product of spatial eigenfunctions $v=v(\s_{ij})$, $\zeta=\zeta(y)$, $\f_k=\f_k(x)$, $k\in\N$, and  temporal eigenfunctions $w=w(t)$; $u$ is evaluated at $\s_{ij}=\chi_{ij}$, where
\begin{equation}
u=wv\zeta\f_k.
\end{equation}
The temporal eigenfunction $w$ is a solution  of the ODE 
\begin{equation}\lae{3.42}
\begin{aligned}
&\frac n{16(n-1)} t^{-m}\frac\pa{\pa t}\big (t^m \pde wt\big )+t^{-2}(\abs\lam^2+\rho^2)w\\
&\qq t^{2-\frac4n}\{(n-1)\abs\xi^2+\bar\mu_k\}w+\frac{n-2}2 t^2\Lam w=0
\end{aligned}
\end{equation}
in $0<t<\un$.
\et

In the next sections we shall solve the ODE and shall also show that for large $n$, $n\ge 17$, and negative $\Lam$ $w$ can be chosen to be an eigenfunction of a self-adjoint operator where the cosmological constant plays the role of an implicit eigenvalue. 

\section{Temporal eigenfunctions: the case $3\le n\le 16$}\las 4
Let us first divide the equation \re{3.42} by $\frac n{16(n-1)}$ to obtain what we consider to be a 
normal form
\begin{equation}\lae{4.1}
\begin{aligned}
&t^{-m}\frac\pa{\pa t}\big (t^m \pde wt\big )+t^{-2}\frac{16(n-1)}n(\abs\lam^2+\rho^2)w\\
&\qq t^{2-\frac4n}
\frac{16(n-1)}n\{(n-1)\abs\xi^2+\bar\mu_k\}w+\frac{16(n-1)}n\frac{n-2}2 t^2\Lam w=0
\end{aligned}
\end{equation}
Using the abbreviations
\begin{equation}\lae{4.2}
\mu_0=\frac{16(n-1)}n(\abs \lam^2+\abs\rho^2),
\end{equation}
\begin{equation}\lae{4.3}
m_1=\frac{16(n-1)}n\{(n-1)\abs\xi^2+\bar\mu_k\}
\end{equation}
and
\begin{equation}\lae{4.4}
m_2=\frac{8(n-1)(n-2)}n 
\end{equation}
we can rewrite the equation \re{4.1} in the form
\begin{equation}\lae{4.5}
\begin{aligned}
t^{-m}\frac\pa{\pa t}\big (t^m \pde wt\big )+t^{-2}\mu_0w
+ t^{2-\frac4n} m_1 w+ t^2 m_2 \Lam w=0.
\end{aligned}
\end{equation}
We shall use two different approaches in solving this ODE depending on the sign of
\begin{equation}\lae{4.6}
\mu_0-\frac {(m-1)^2}4.
\end{equation}
Let us recall that
\begin{equation}
m=\frac{(n-1)(n+2)}2.
\end{equation}
and
\begin{equation}
\rho^2=\frac{(n-1)^2n}{12}.
\end{equation}
One can easily check that
\begin{equation}
\frac{16(n-1)}n\rho^2-\frac{(m-1)^2}4=
\begin{cases}
>1,&3\le n\le 16,\\
<-238,&17\le n.
\end{cases}
\end{equation}
Hence, in case $3\le n\le 16$ the term in \re{4.6} will be strictly larger than $1$ for all values of $\abs\lam$ and in case $n\ge 17$ strictly negative for small values of $\abs\lam$, or more precisely, for all
\begin{equation}\lae{4.10}
\frac{16(n-1)}n\abs\lam^2<238.
\end{equation}

Let us first consider the case $3\le n\le 16$ and let us rewrite equation \re{4.5} in the form
\begin{equation}
\Ddot w+m t^{-1}w+ t^{-2}\{\mu_0+ m_2 t^{{4-\frac4n}}+m_3\Lam  t^4 \}w=0\qq\A \,t>0.
\end{equation}
Then we look at the more general equation
\begin{equation}\lae{1.9}
\Ddot w+m t^{-1}w+ t^{-2}(\mu_0+q_0(t))w=0\qq\A\, t>0,
\end{equation}
for which we proved in  \cite[Theorem 1.1]{cg:bbang} the following theorem
\bt\lat{4.1}
Let us assume that the constants $m,\m_0$ and the real function $q_0\in C^1({\R[]}_+)$ have the properties
\begin{equation}
m>1,
\end{equation}
\begin{equation}
1<\mu_0-\frac{(m-1)^2}4\equiv 1+\ga,\qq \ga >0, 
\end{equation}
and
\begin{equation}
\lim_{t\ra 0}q_0(t)=0.
\end{equation}
Then any non-trivial solution $w$ of \re{1.9} satisfies
\begin{equation}
\lim_{t\ra 0}(\abs w^2+t^2\abs{\dot w}^2)=\un
\end{equation}
as well as
\begin{equation}
\limsup_{t\ra 0}\abs w^2=\un.
\end{equation}
\et
We also described the oscillation behaviour of $w$ near $t=0$, which can be considered to be a big bang of the solutions, as to be asymptotically equal to the oscillations of the solutions of the ODE
\begin{equation}
\Ddot w+ mt^{-1} w+\mu_0  t^{-2}w=0\qq\A\,t>0,
\end{equation}
\cf \cite[Theorem 3.2]{cg:bbang}.  The solutions of the above equation are 
\begin{equation}
w(t)=t^{-\frac{(m-1)}2}e^{i\mu\log t},\qq \mu>0,
\end{equation}
where
\begin{equation}
\mu^2=\m_0-\frac {(m-1)^2}4,
\end{equation}
see \cite[equ. (273)]{cg:qgravity3}. 

\section{Temporal eigenfunctions: the case $n\ge 17$}\las 5

\subsection{Treating $\Lam$ as an eigenvalue}\lass{5.1}
Now, let us consider the case $n\ge 17$ assuming in addition that \fre{4.10} is  satisfied such that
\begin{equation}\lae{5.1}
\bar\mu\equiv \mu_0-\frac {(m-1)^2}4<0
\end{equation}
and also that 
\begin{equation}\lae{5.2}
\Lam<0.
\end{equation}
The last two assumptions shall allow us to consider \fre{4.5}  as an implicit eigenvalue equation where $\Lam$ plays the role of the eigenvalue. We shall prove that the corresponding operator is self-adjoint with a pure point spectrum provided the constant $m_1$ in \re{4.5}, which is defined by \re{4.3}, is strictly positive. This can easily be arranged by choosing $\abs\xi$ large enough. Notice also that at most finitely many eigenvalues $\bar\mu_k$ are negative.

The equation \re{4.5} can be written in  the equivalent form
\begin{equation}\lae{5.3}
-t^{-m}\frac\pa{\pa t}\big (t^m \pde wt\big )-t^{-2}\mu_0w
- t^2 m_2 \Lam w=t^{2-\frac4n} m_1 w\qq \A\,t>0.
\end{equation}
We have a similar equation, or, since the constants, $m_1, m_2$, are not specified and their actual positive values are irrelevant, an identical equation already solved by spectral analysis in \cite[Section 4 \& Section 6]{cg:qfriedman}. Therefore, we shall only outline the proof by giving the necessary definitions and stating the results but referring the actual proofs to the old paper.

Closely related to equation \re{5.3} is the following equation
\begin{equation}\lae{5.4}
-t^{-1}\frac\pa{\pa t}\big (t \pde ut\big )-t^{-2}\bar\mu u
- t^2 m_2 \Lam u=t^{2-\frac4n} m_1 u\qq \A\,t>0,
\end{equation}
where $\bar\mu$ is defined in \re{5.1}.  If $w\in C^2(\R[*]_+)$ is a solution of \re{5.3} then
\begin{equation}
u=t^{\frac{m-1}2} w
\end{equation}
is a solution of \re{5.4} and vice versa, as can be easily checked. The operator
\begin{equation}\lae{5.6}
B u=-t^{-1}\frac\pa{\pa t}\big (t \pde ut\big )-t^{-2}\bar\mu u
\end{equation}
is known as a Bessel operator.

\bd
Let $I=(0,\un)$  and let $q\in\R[]$. Then we define
\begin{equation}
L^2(I,q)=\set{u\in L^2_{\tup{loc}}(I,\R[])}{\int_Ir^q\abs u^2<\un}.
\end{equation}
$L^2(I,q)$ is a Hilbert space with scalar product 
\begin{equation}
\spd{u_1}{u_2}_q=\int_I r^qu_1u_2,
\end{equation}
but let us emphasize that we shall apply this definition only for $q\not=2$. The scalar product
$\spd\cdot\cdot_2$ will be defined differently.
\ed
We consider real valued functions for simplicity but we could just as well allow complex valued functions with the standard scalar product, or more precisely, sesquilinear form.
\bd\lad{5.2}
For functions $w, u\in C^\un_c(I)$ define the operators
\begin{equation}
\hat A_1w=-t^{-m}\frac\pa{\pa t}\big (t^m \pde wt\big )-t^{-2}\mu_0w
- t^2 m_2 \Lam w
\end{equation}
and
\begin{equation}
A_1u=-t^{-1}\frac\pa{\pa t}\big (t \pde ut\big )-t^{-2}\bar\mu u
- t^2 m_2 \Lam u,
\end{equation}
as well as the scalar product 
\begin{equation}\lae{5.12}
\spd{u_1}{u_2}_2=\spd{Bu_1+t^2m_2u_1}{u_2}_1\qq\A\, u_1,u_2\in C^\un_c(I).
\end{equation}
\ed
The right-hand side of \re{5.12} is an integral. Integrating by parts we deduce
\begin{equation}\lae{5.12.1}
\spd{u_1}{u_2}_2=\int_I(t\dot u_1\dot u_2-\bar\mu t^{-1}u_1u_2+t^3m_2 u_1u_2),
\end{equation} 
i.e., the scalar product is indeed positive definite because of the assumption \re{5.1}. Let us define the norm
\begin{equation}\lae{5.13}
\norm u_2^2=\spd uu_2\qq \A\, u\in C^\un_c(I)
\end{equation}
and the Hilbert space $\mc H_2=\mc H_2(I)$ as the closure of $C^\un_c(I)$ with respect to the norm $\norm\cdot_2$.
\bpp
The functions $u\in \mc H_2$ have the properties
\begin{equation}\lae{5.15}
u\in C^0([0,\un)),
\end{equation}
\begin{equation}\lae{5.16}
\abs {u(t)}\le c\norm u_2\qq\A\,t\in I,
\end{equation}
where $c=c(\bar\mu, m_2,\abs\Lam)$,
\begin{equation}\lae{5.17}
\lim_{t\ra 0} u(t)=0
\end{equation}
and
\begin{equation}\lae{5.18}
\abs{u(t)}\le c \norm u_2 t^{-1}\qq\A\, t\in I,
\end{equation}
where $c$ is a different constant depending on $\bar\mu, m_2$ and $\abs\Lam$.
\epp
\bp
Let us first assume $u\in C^\un_c(I)$ and let $\de>0$, then
\begin{equation}\lae{5.19}
u^2(\de)=2\int_0^\de\dot u u\le\int_0^\de t\abs{\dot u}^2+\int_0^\de t^{-1}\abs u^2.
\end{equation}
This estimate is  also valid for any $u\in \mc H_2$ by approximation which in turn implies the relations \re{5.16}, \re{5.17} and also \re{5.15} since $u$ is certainly continuous in $I$.

It remains to prove \re{5.18}. Let $u\in \mc H_2$ and define $v=v(\tau)$ by
\begin{equation}
v(\tau)=u(\tau^{-1}),
\end{equation}
where $\tau=t^{-1}$ for all $t>0$. Applying simple calculus arguments we then obtain
\begin{equation}\lae{5.21}
\begin{aligned}
\int_0^\un\{\tau\abs{v'}^2-\bar\mu\tau^{-1}\abs v^2+m_2 \tau^{-5}\abs v^2\}d\tau=\norm u_2^2
\end{aligned}
\end{equation}
as well as
\begin{equation}\lae{5.22}
\begin{aligned}
\int_0^\un\{\tau\abs{v'}^2-\bar\mu\tau^{-1}\abs v^2\}d\tau=\int_0^\un\{t\abs{\dot u}^2-\bar\mu t^{-1}\abs u^2\}dt.
\end{aligned}
\end{equation}
Moreover, first assuming, as before, that $u$ and hence $v$ are test functions we argue as in \re{5.19} that for any $\de>0$
\begin{equation}
\begin{aligned}
v^2(\de)=2\int_0^\de v'v&\le 2\bigg(\int_0^\de \tau \abs{v'}^2\bigg)^\frac12\bigg(\int_0^\de \tau^{-1} \abs{v}^2\bigg)^\frac12\\
&\le  2\bigg(\int_0^\de \tau \abs{v'}^2\bigg)^\frac12\bigg(\int_0^\de \tau^{-5} \abs{v}^2\bigg)^\frac12 \de^2\\
&\le c \norm u_2^2\de^2,
\end{aligned}
\end{equation}
where we used \re{5.21} for the last inequality and where $c=c(\bar\mu,m_2)$. Setting $\de=t^{-1}$ for arbitrary $t>0$ we have proved the estimate \re{5.18} for test functions and hence for arbitrary $u\in \mc H_2$. 
\ep

We are now ready to solve the equation \re{5.4} as an implicit  eigenvalue equation. First, we need

\bl\lal{5.4}
Let $K$ be the quadratic form
\begin{equation}\lae{5.23} 
K(u)=m_1\int_I t^{3-\frac {4}n} u^2,
\end{equation}
then $K$ is compact in $\mc H_2$, i.e., 
\begin{equation}
u_i\whc{\mc H_2} u\q\im\q K(u_i)\ra K(u), 
\end{equation}
and positive definite, i.e.,
\begin{equation} 
K(u)>0\qq\A\,u\ne 0.
\end{equation}
\el
For a proof we refer to \cite[Lemma 6.8]{cg:qfriedman}. Then, we look at the eigenvalue problem for $u\in \mc H_2$
\begin{equation}\lae{5.26}
Bu+m_2 t^2 u=\lam m_1 t^{2-\frac 4n}u,
\end{equation}
or equivalently,
\begin{equation}\lae{5.27}
\tilde B(u,v)\equiv \spd{Bu+m_2 t^{2}u}v_1=\lam  K(u,v)\q\A\, v\in\mc H_2,
\end{equation}
where $K(u,v)$ is the bilinear form associated with $K$.
\bt\lat{5.5}
The eigenvalue problem \re{5.27} has countably many solutions $(\lam_i,\tilde u_i)$, $\tilde u_i\in \mc H_2$, with the properties
\begin{equation}
\lam_i<\lam_{i+1}\qq\A\,i\in\N,
\end{equation}
\begin{equation}
\lim_i\lam_i=\un,
\end{equation}
\begin{equation}
K(\tilde u_i,\tilde u_j)=\de_{ij}.
\end{equation}
 The pairs $(\lam_i, \tilde u_i)$ are  recursively defined  by the variational problems
 \begin{equation}\lae{5.31}
\lam_0=\tilde B(\tilde u_0)=\inf \bigg\{\frac{\tilde B(u)}{K(u)}:0\not=u\in \mc H_2\bigg\}
\end{equation}
and for $i>0$
 \begin{equation}
\lam_i=\tilde B(\tilde u_i)=\inf \bigg\{\frac{\tilde B(u)}{K(u)}:0\not=u\in \mc H_2,\, K(u,u_j)=0, \, 0\le j\le i-1\bigg\}.
\end{equation}
 The $(\tilde u_i)$ form a Hilbert space basis in $\mc H_2$ and in $L^2(I,3-\frac4n)$, the eigenvalues are strictly  positive and the eigenspaces are one dimensional. 
\et
\bp
This theorem is well-known and goes back to the book of Courant-Hilbert \cite{courant-hilbert-I}, though in a general separable Hilbert space the eigenvalues are not all positive and the eigenspaces are only finite dimensional . For a proof in the general case we refer to \cite[Theorem 1.6.3, p. 37]{cg:pdeII}. 

The positivity of the eigenvalues in the above theorem is obvious and the fact that the eigenspaces are one dimensional is proved by contradiction. Thus, suppose there exist an eigenvalue $\lam=\lam_i$ and two corresponding linearly independent eigenfunctions $u_1,u_2\in \mc H_2$. Then, for any $t_0>0$ there would exist an eigenfunction $u\in \mc H_2$ with eigenvalue $\lam$  satisfying $v(t_0)=0$ and the equation \re{5.26}. Multiplying this equation by $u$ and integrating the result in the interval $(0,t_0)$ with respect to the measure $t\, dt$ we obtain
\begin{equation}
\int_0^{t_0}-\bar\mu t^{-1} u^2\le t_0^{4-\frac4n}\int_0^{t_0}\lam m_1 t^{-1} u^2,
\end{equation}
where we used
\begin{equation}
1\le \frac{t_0}t,\qq \A\,t\in (0,t_0),
\end{equation}
yielding a contradiction if $t_0$ is sufficiently small.
\ep

The functions
\begin{equation}\lae{5.33}
u_i(t)=\tilde u_i(\lam_i^{-\frac n{4(n-1)}} t)
\end{equation}
then satisfy the equation 
\begin{equation}
Bu_i+m_2\lam_i^{-\frac{n}{n-1}}t^{2}u_i=m_1 t^{2-\frac{4}n} u_i
\end{equation}
and they are mutually orthogonal with respect to the bilinear form
\begin{equation}
\int_It^3 uv,
\end{equation}
as one easily checks. Furthermore,  the following lemma is valid:
\bl
Let $(\lam,u) \in \R[*]_+\times \mc H_2$,  be a solution of
\begin{equation}
Bu+m_2\lam^{-\frac{n}{n-1}}t^{2}u=m_1 t^{2-\frac4 n} u, 
\end{equation}
then there exists $i$ such that 
\begin{equation}
\lam=\lam_i\q\wed\q u\in\langle u_i\rangle.
\end{equation}
\el
\bp
Define
\begin{equation}
\tilde u(t)=u(\lam ^{\frac n{4(n-1)}}t),
\end{equation}
then the pair $(\lam,\tilde u)$ is a solution of the equation \re{5.26}, hence the result.
\ep

Thus we have proved
\bt\lat{5.7}
There  are countably many solutions $(\Lam_i,u_i)$ of the implicit eigenvalue problem
\begin{equation}\lae{5.39}
Bu_i-m_2\Lam_it^{2}u_i=m_1 t^{2-\frac{4}n} u_i
\end{equation}
with eigenfunctions $u_i\in \mc H_2$ such that
\begin{equation}
\Lam_i<\Lam_{i+1}<0\qq\A\,i\in\N,
\end{equation}
\begin{equation}
\lim_i\Lam_i=0,
\end{equation}
and their multiplicities are one. 
The transformed eigenfunctions 
\begin{equation}
\tilde u_i(t)=u_i(\lam_i^{\frac{n}{4(n-1)}}t),
\end{equation}
where
\begin{equation}
\lam_i=(-\Lam_i)^{-\frac {n-1}{n}},
\end{equation}
form a basis of $\mc H_2$ and also of $L^2(I,1)$.
\et
\br\lar{5.8}
The eigenfunctions $\tilde u_0$ \resp  $u_0$ corresponding to the smallest eigenvalues $\lam_0$ \resp $\Lam_0$ do not change sign in $I$, since
\begin{equation}
\tilde B(\abs u)\le \tilde B(u)\qq\A\, u\in \mc H_2,
\end{equation}
in view of \re{5.6}, and hence we deduce that $\abs{\tilde u_0}$ is also an eigenfunction with eigenvalue $\lam_0$, i.e., we may assume $\tilde u_0\ge 0$. But if $\tilde u_0$ would vanish in a $t_0>0$ then its derivative $\tilde u'_0$ would also vanish in $t_0$ yielding $\tilde u_0$ would completely vanish, a contradiction.
\er

In \rd{5.2} we defined the operators $A_1$ and $\hat A_1$. The operator $A_1$ can be expressed with the help of the Bessel operator $B$ as
\begin{equation}
A_1u=Bu-t^2 m_2\Lam u.
\end{equation}
Let us express $\hat A_1$ similarly as
\begin{equation}
\hat A_1w=\hat Bw-t^2 m_2\Lam w,
\end{equation}
where
\begin{equation}
\hat Bw=-t^{-m}\frac\pa{\pa t}\big (t^m \pde wt\big )-t^{-2}\mu_0w.
\end{equation}
We claim that $B$ and $\hat B$ are unitarily equivalent.
\bl\lal{5.8}
Let $\f$ be the linear map from $L^2(I,m)$ to $L^2(I,1)$ defined by
\begin{equation}
\f(w)=u=t^{\frac{m-1}2}w.
\end{equation}
Then $\f$ is unitary and, if $B$ \resp $\hat B$ are defined in $C^\un_c(I)$, the relation
\begin{equation}\lae{3.10}
\hat B=\f^{-1}\circ B\circ \f
\end{equation}
is valid. 
\el
Since we assume for simplicity the Hilbert spaces to be real Hilbert spaces it would be better to call the map $\f$ \tit{orthogonal} but the result would be same if we would consider complex valued functions and the corresponding scalar products. 

For the simple proof of the lemma we refer to \cite[Lemma 4.1]{cg:qfriedman}. Moreover, for any measurable function $f=f(t)$ we have
\begin{equation}
\spd{f\f (w)}{\f (v)}_1=\spd{fw}{v}_m\qq\A\, v,w\in C^\un_c(I).
\end{equation}
Hence, we infer
\begin{equation}\lae{5.51}
\begin{aligned}
\spd {\f(w)}{\f(v)}_2&=\spd{B\f(w)+t^2m_2\f(w)}{\f(v)}_1\\
&=\spd{\hat Bw+t^2m_2w}v_m\qq\qq\qq\A\, v,w\in C^\un_c(I),
\end{aligned}
\end{equation}
\begin{equation}
\spd{\hat Bw}v_m=\spd{B\f(w)}{\f(v)}_1
\end{equation}
and we deduce,  by setting $u=\f(w)=t^\frac{m-1}2w$, that 
\begin{equation}
\spd{\hat Bw}w_m=\spd{Bu}u_1=\int_I(t\abs{\dot u}^2-\bar\mu t^{-1} \abs u^2)>0,
\end{equation}
or equivalently,
\begin{equation}
\int_It^{m}\abs{\dot w}^2=\int_I(t\abs{\dot u}^2-\bar\mu t^{-1} \abs u^2)+\mu_0\int_It^{m-2}\abs w^2\qq\A\, w\in C^\un_c(I).
\end{equation}
Let us recall that $\bar\mu<0$ and $\mu_0>0$. 

\br\lar{5.9}
Defining  the Hilbert space $\mc{\hat H}_2$ by
\begin{equation}\lae{5.55}
\mc{\hat H}_2=\set{w=t^{-\frac{m-1}2}u}{u\in {\mc H}_2} 
\end{equation}
with norm  
\begin{equation}
\nnorm w_2=\norm u_2
\end{equation}
and the quadratic form $\hat K$ by
\begin{equation}\lae{5.57} 
\hat K(w)=\spd{m_1t^{2-\frac4n}w}w_m=\spd{m_1t^{2-\frac4n} u}u_1=K(u)\qq\A\,w\in \mc{\hat H}_2
\end{equation}
it is fairly easy to verify that all results in \rt{5.5} remain valid if $B, \tilde B, K, {\mc H}_2$ are replaced by $\hat B, \tilde{\hat B},\hat K,\mc{\hat H}_2$. The eigenvalues $\lam_i$ are identical and the eigenfunctions are related by
\begin{equation}
\tilde w_i=t^{-\frac{m-1}2}\tilde u_i,
\end{equation}
i.e.,
\begin{equation}\lae{5.59}
\hat B\tilde w_i+t^2m_2t^2\tilde w_i=\lam_im_1t^{2-\frac4n}\tilde w_i.
\end{equation}
Similarly, the transformed eigenfunctions $u_i$ in \rt{5.7} correspond to 
\begin{equation}
w_i=t^{-\frac{m-1}2}u_i
\end{equation}
satisfying
\begin{equation}\lae{5.61}
\hat Bw_i-m_2\Lam_it^{2}u_i=m_1 t^{2-\frac{4}n} w_i,
\end{equation}
which is the original ODE \fre{4.5} with $\Lam=\Lam_i$.
\er
For completeness let us restate \rt{5.7} in the new setting
\bt\lat{5.10}
There  are countably many solutions $(\Lam_i,w_i)$ of the implicit eigenvalue problem
\begin{equation}
\hat Bw_i-m_2\Lam_it^{2}w_i=m_1 t^{2-\frac{4}n} w_i
\end{equation}
with eigenfunctions $w_i\in \mc{\hat H}_2$ such that
\begin{equation}
\Lam_i<\Lam_{i+1}<0\qq\A\,i\in\N,
\end{equation}
\begin{equation}
\lim_i\Lam_i=0,
\end{equation}
and their multiplicities are one.  
The transformed eigenfunctions 
\begin{equation}
\tilde w_i(t)=w_i(\lam_i^{\frac{n}{4(n-1)}}t),
\end{equation}
where
\begin{equation}
\lam_i=(-\Lam_i)^{-\frac {n-1}{n}},
\end{equation}
form a basis of $\mc{\hat H}_2$ and also of $L^2(I,m)$.
\et

Finally, let us show how the eigenvalue equations \re{5.26} \resp \re{5.59} can be considered to be eigenvalue equations of an essentially self-adjoint operator in an appropriate Hilbert space. We shall first demonstrate it for  the equation \re{5.26}.

Let $\f_0(t)$ be defined by
\begin{equation}\lae{5.69}
\f_0(t)=m_1t^{3-\frac4n}\qq\A\, t\in I
\end{equation}
and define the Hilbert space $\mc H$ as $L^2(I,d\mu)$ with respect to the measure
\begin{equation}\lae{5.70.1}
d\mu=\f_0 dt.
\end{equation}
Moreover, denote the scalar product in $\mc H$ by $\spd\cdot\cdot$ and the corresponding norm by $\norm\cdot$. Note that, in view of \re{5.23},
\begin{equation}
\spd uv=K(u,v).
\end{equation}
The operator
\begin{equation}\lae{5.71.1} 
Au=\f_0^{-1}\big\{-(\frac\pa{\pa t}\big (t \pde ut\big )-t^{-1}\bar\mu u+t^3 m_2  u\big\}\qq\A\, u\in C^\un_c(I)
\end{equation}
is densely defined and symmetric in $\mc H$ such that  
\begin{equation}
\spd {Au}v=\spd uv_2\qq\A\, u,v\in C^\un_c(I)
\end{equation}
The above relation is also valid for all $u,v\in {\mc H}_2$  by partial integration. Hence the domain $D( A$) of $A$ is contained in ${\mc H}_2$.  In view of equation \re{5.26} we infer
\begin{equation}\lae{5.70}
A\tilde u_i=\lam_i\tilde u_i,\qq\A\, i\in \N,
\end{equation}
i.e., $\tilde u_i$ is an eigenfunction of $A$ in the classical sense. Since $A$ is symmetric $A$ is closable. Let $\bar A$ be the closure of $A$. If $\bar A$ is surjective
\begin{equation}\lae{5.71}
R(\bar A)=\mc H,
\end{equation}
then $\bar A$ is self-adjoint and $A$ essentially self-adjoint. These are well-known facts. Let us prove \re{5.71} for convenience.
\bl
$\bar A$ is surjective. 
\el
\bp
First we observe that $R(A)$  is dense in $\mc H$ because of \re{5.70}. Indeed the eigenfunctions $(\tilde u_i),i\in\N,$ are complete and the eigenvalues are strictly positive, \cf \rt{5.5}.

Next, let $v\in \mc H$ be arbitrary and let $u_i\in D(A)$ be a sequence such that   
\begin{equation}
Au_i\ra v, 
\end{equation}
then
\begin{equation}
\begin{aligned}
\lam_0\norm {u_i-u_j}^2&=\lam_0\spd {u_i-u_j}{u_i-u_j}\le \spd{A(u_i-u_i)}{u_i-u_j}\\
&\le \norm{A(u_i-u_j)}\norm{u_i-u_j},
\end{aligned}
\end{equation}
where $0<\lam_0$ is the smallest eigenvalue, \cf \re{5.31}. Hence
\begin{equation}
\lam_0\norm {u_i-u_j}\le \norm{A(u_i-u_j)},
\end{equation}
i.e., $(u_i)$ is a Cauchy sequence which implies  $v\in R(\bar A)$, completing the proof of the lemma.
\ep

In case of equation \re{5.59} we define $\hat\f_0(t)$ by
\begin{equation}
\hat\f_0(t)=m_1t^{m+2-\frac4n}\qq\A\, t\in I
\end{equation}
and define the Hilbert space $\mc {\hat H}$ as $L^2(I,d\hat\mu)$ with respect to the measure
\begin{equation}\lae{5.80}
d\hat\mu=\hat\f_0 dt.
\end{equation}
Moreover, denote the scalar product in $\mc {\hat H}$ by $\spdd\cdot\cdot$ and the corresponding norm by $\nnorm\cdot$. Note that, in view of \re{5.57}, 
\begin{equation}
\spdd wv=\hat K(w,v)
\end{equation}
The operator
\begin{equation}\lae{5.81.1}
\hat Aw=\hat\f_0^{-1}\big\{-(\frac\pa{\pa t}\big (t^m \pde wt\big )-t^{m-2}\mu_0 w+t^{m+2} m_2  w\big\}\qq\A\, w\in C^\un_c(I)
\end{equation}
is densely defined and symmetric in $\mc {\hat H}$ such that
\begin{equation}\lae{5.81}
\spdd {\hat Aw_1}{w_2}=\spd {Au_1}{u_2}=\spd {u_1}{u_2}_2\qq\A\, w_1,w_2\in C^\un_c(I),
\end{equation}
where 
\begin{equation}
u_i=\f(w_i)\equiv t^\frac{m-1}2w_i,\qq i=1,2,
\end{equation}
\cf the definition of $\f$ in \rl{5.8} and also the equation \re{5.51}. If equation \re{5.81} would be valid for all $w_1,w_2\in D(\hat A)$ then $\hat A$ and $A$ would be unitarily equivalent, since $\f$ is obviously a unitary (orthogonal) map between $\mc{\hat H}$ and $\mc H$.  

This is indeed the case as one can easily infer from \rr{5.9}, hence
\begin{equation}\lae{5.85}
\hat A \tilde w_i=\lam _i \tilde w_i,
\end{equation}
where 
\begin{equation}
\tilde w_i=t^{-\frac{m-1}2}\tilde u_i
\end{equation}
and $\tilde u_i$ an eigenfunction $A$ with eigenvalue $\lam_i$. The domain of $\hat A$ satisfies
\begin{equation}
D(\hat A)=\f^{-1}(D(A)).
\end{equation}
\setcounter{subsection}{1}
\subsection{Treating $\Lam$ as a fixed cosmological constant} \lass{5.2}

If we want to define a partition function and entropy for our quantum system we have to consider  $\Lam$ to be a fixed cosmological constant and not a parameter which can also play the role of an implicit eigenvalue. Our approach to solve the ODE \fre{4.5} then is similar but different.
\setcounter{subsection}{13 } 
First, let us express equation \re{4.5} in the equivalent form 
\begin{equation}\lae{5.88}
\begin{aligned}
\hat\f_0^{-1}\bigg\{&-\frac\pa{\pa t}\big (t^m \pde wt\big )-t^{m-2}\mu_0w
- t^{m+2} m_2 \Lam w\bigg\}\\
&\qq\qq- \frac{16(n-1)}n\{(n-1)\abs\xi^2+\bar\mu_k\}  w=0,
\end{aligned}
\end{equation}
where
\begin{equation}
\hat\f_0(t)=t^{m+2-\frac4n}
\end{equation}
and where we used the definition \fre{4.3} of $m_1$. The term
\begin{equation}
(n-1)\abs\xi^2+\bar\mu_k
\end{equation}
is an eigenvalue of the operator in \fre{3.34}. $\abs \xi^2$ with $\xi\in \R[n_1]$ is a continuous eigenvalue while the sequence $\bar\mu_k$, $k\in\N$, satisfies the relations \re{3.39} and \re{3.40}. The operator 
\begin{equation}\lae{5.91}
\hat H_0w\equiv \hat\f_0^{-1}\bigg\{-\frac\pa{\pa t}\big (t^m \pde wt\big )-t^{m-2}\mu_0w
- t^{m+2} m_2 \Lam w\bigg\}
\end{equation}
is identical to the operator $\hat A$ defined in \re{5.81.1} if $\Lam=-1$. The properties we proved for $\hat A$ are also valid for $\hat H_0$ by simply replacing  $-m_2\Lam$ by a positive constant $m_2'$. Thus, we know that $\hat H_0$ is essentially self-adjoint in the Hilbert space $\mc{\hat H}=L^2(I,d\hat\mu)$, \cf \re{5.80} with a complete system of eigenfunctions $w_i$, $i\in\N$, and corresponding eigenvalues 
\begin{equation}\lae{5.92}
0<\lam_0<\lam_1<\lam_2<\cdots
\end{equation}
The eigenspaces are all one dimensional and the ground state $w_0$ does not change sign, \cf \frr{5.8}. 

Note that we denote the eigenfunctions by $w_i$ and not by $\tilde w_i$ since they will not be transformed to obtain the final solutions of the ODE. Instead they will be the solutions of the ODE satisfying 
\begin{equation}\lae{5.93}
\hat H_0w_i=\lam_i w_i\qq\A\,i\in\N.
\end{equation}
But  $w_i$ is a solution of  the ODE \re{5.88} if and only if there exist $j$ and $\xi$ such that 
\begin{equation}\lae{5.94}
\lam_i=\frac{16(n-1)}n\{(n-1)\abs\xi^2+\bar\mu_j\}
\end{equation}
Obviously, the previous equation can only be satisfied for all  $\lam_i$ iff
\begin{equation}\lae{5.95}
\lam_0\ge \frac{16(n-1)}n\bar \mu_0.
\end{equation}
In \cite[Lemma 6.4.9, p. 172]{cg:qgravity-book} we proved the following lemma:
\bl\lal{5.15}
Let $\lam_i$ be the temporal eigenvalues depending on $\Lam<0$ and let $\bar\lam_i$ be the corresponding eigenvalues for
\begin{equation}
\Lam=-1,
\end{equation}
then
\begin{equation}\lae{5.97}
\lam_i=\bar\lam_i\abs\Lam^\frac{n-1}n.
\end{equation}
\el
Thus, we deduce
\bc\lac{5.16}
Suppose that $\bar\m_0>0$ and define  $\Lam_0<0$ by
\begin{equation}
\abs{\Lam_0}^\frac{n-1}n=\bar\lam_0^{-1} \frac{16(n-1)}n\bar \mu_0,
\end{equation}
then, the inequality \re{5.95} is satisfied provided
\begin{equation}\lae{5.99}
\abs\Lam\ge \abs{\Lam_0}.
\end{equation}
\ec
The inequality \re{5.95} is  always satisfied if $\bar\mu_0\le 0$.

The eigenvalues on the right-hand side of equation \re{5.94}, i.e., the sum inside the braces, are the eigenvalues of the operator defined in \fre{3.34} which can be written as the sum
\begin{equation}\lae{3.34.1}
-(n-1)\D_\de+A,
\end{equation}
where $A$ is a uniformly elliptic operator on a compact Riemannian manifold, \cf equation \fre{3.38}. Hence, we can interpret the right-hand side of \re{5.94} as  eigenvalues of the operator
\begin{equation}\lae{5.101}
H_1=-\frac{16(n-1)^2}n\D_\de+\frac{16(n-1)}nA.
\end{equation}
To facilitate a comparison with former results in \cite[Sections 6.4 \& 6.5]{cg:qgravity-book} let us define
\begin{equation}
\tilde A=\frac{16(n-1)}nA
\end{equation}
and  
\begin{equation}
\tilde\mu_j=\frac{16(n-1)}n\bar\mu_j,
\end{equation}
then $\tilde A$ has the same eigenfunctions as $A$ with eigenvalues $\tilde\mu_j$ instead of $\bar\mu_j$ and the condition \re{5.94} can be rephrased in the form 
\begin{equation}\lae{5.104}
\lam_i=\frac{16(n-1)^2}n\abs\xi^2+\tilde\mu_j
\end{equation}
and the inequality \re{5.95} can now be expressed as
\begin{equation}\lae{5.105}
\lam_0\ge \tilde\mu_0.
\end{equation}
In \cite[equ. (6.4.67), p.166]{cg:qgravity-book} we considered an operator $H_1$ which was similarly defined as the operator in \re{5.101}, the only difference was that the Laplacian $\D_\de$ was defined in $\R[]$, i.e., the dimension $n_1$ was equal to one. In this case it is fairly simple to determine the tempered eigendistributions  $\zeta_{ijk}$  in $\msc S'(\R[])$ satisfying
\begin{equation}
-\zeta''_{ijk}=\om_{ij}^2\zeta_{ijk},\qq k=1,2,
\end{equation}
where
\begin{equation}
\zeta_{ij1}(\tau)=\frac1{\sqrt{2\pi}}e^{i\om_{ij}\tau}
\end{equation}
and
\begin{equation}
\zeta_{ij2}(\tau)=\frac1{\sqrt{2\pi}}e^{-i\om_{ij}\tau},
\end{equation}
where
\begin{equation}
\om_{ij}\ge 0
\end{equation}
is defined by the relation
\begin{equation}\lae{5.110}
\lam_i=\tilde \mu_j+\frac{16(n-1)^2}n\om_{ij}^2.
\end{equation}
In the higher dimensional case, $n_1>1$, we have a whole continuum of vectors $\xi\in\R[n_1]$
satisfying \re{5.104}, and hence, a whole continuum of eigendistributions which we cannot handle---neither physically nor mathematically. Therefore, let us pick a finite numbers of unit vectors $\xi_k\in\R[n_1]$, $1\le k\le k_1$ which are fixed. Then the eigendistributions are defined by
\begin{equation}\lae{5.111}
\zeta_{ijk}(y)=(2\pi)^{-\frac{n_1}2}e^{i\om_{ij}\spd{\xi_k}y},\qq\, 1\le k\le k_1,
\end{equation}
where
\begin{equation}
\lam_i=\tilde \mu_j+\frac{16(n-1)^2}n\om_{ij}^2
\end{equation}
if $\tilde\mu_j<\lam_i$.
We consider the eigendistributions $\zeta_{ijk}$ to be mutually orthogonal since their Fourier transforms
\begin{equation}
\hat\zeta_{ijk}=\de_{\om_{ij}\xi},
\end{equation}
which are Dirac measurers, have disjoint supports. 

Now, we are able to define the eigenfunctions of the operator $H_1$ in \re{5.101}.
\bd
Let $\f_j\in L^2(M)$ be the mutually orthogonal  unit eigenvectors of $\tilde A$  with corresponding eigenvalues $\tilde\mu_j$ and assume either that $\bar\mu_0\le 0$ or that $\Lam$ satisfies the condition \re{5.99} in \rc{5.16}. Then, for any eigenvalue $\lam_i$, we define
\begin{equation}
N_i=\{j\in\N:\tmu_j\le\lam_i\}
\end{equation}
and $\om_{ijk}\ge 0$ such that
\begin{equation}
\frac{16(n-1)^2}n\om_{ijk}^2+\tmu_j=\lam_i,\qq\, 1\le k\le k_1,
\end{equation}
provided $\tilde\mu_j<\lam_i$. If $\tilde\mu_j=\lam_i$, then we choose $\om_{ijk}=0$ and the multiplicity will be only the multiplicity of $\tilde\mu_j$. 

Note that
\begin{equation}
0\in N_i\qq\A\, i\in\N,
\end{equation}
since 
\begin{equation}
\tmu_0\le \tilde\lam_0,
\end{equation}
For $j\in N_i$ define the eigenfunctions $v_{ijk}$ of $H_1$ by
\begin{equation}\lae{5.118}
v_{ijk}=\zeta_{ijk}\f_j,
\end{equation}
where this distinction only occurs if
\begin{equation}
\tilde\mu_j<\lam_i,
\end{equation}
such that
\begin{equation}\lae{5.120}
H_1v_{ijk}=\lam_i v_{ijk}.
\end{equation}
\ed

\br\lar{5.17}
$H_1$ has the same eigenvalues $\lam_i$ as $\hat H_0$ but with finite multiplicities $m(\lam_i)$ in general different from one which can be estimated from above by 
\begin{equation}
m(\lam_i)\le  k_1 \card N_i \equiv k_1 n(\lam_i).
\end{equation}
Recall that we labelled the eigenvalues $\tilde\mu_j$ by including their multiplicities, \cf \fre{3.39}. Hence, if 
\begin{equation}
\tilde\mu_j<\lam_i\qq\A\, j\in N_i
\end{equation}
then
\begin{equation}
m(\lam_i)=k_1 n(\lam_i).
\end{equation}
\er
Let us now define a separable Hilbert space $\mc H$ such that $H_1$ is essentially self-adjoint in $\mc H$ and its eigenvectors with eigenvalues $\lam_i$ form an ONB, an orthonormal basis.

First we declare the countable eigenvectors in \re{5.120} to be mutually orthogonal unit vectors and we consider them to be the Hamel basis of the complex vector space $\mc H'$. Since the basis vectors are mutually orthogonal unit vectors they also define a unique hermitian scalar product in $\mc H'$. Let $\mc H$ be the completion of $\mc H'$ with respect to that scalar product. Since the eigenvalues $\lam_i$ are positive and bounded from below by $\lam_0$, we could proved in \cite[Lemma 6.5.1, p. 174]{cg:qgravity-book} the following lemma:
\bl\lal{5.18}
The linear operator $H_1$ with domain $\mc H'$ is essentially self-adjoint  in $\mc H$. Let $\bar H_1$ be its closure, then the only eigenvectors of $\bar H_1$ are those of $H_1$.
\el
\br
In the following we shall write $H_1$ instead of $\bar H_1$ and we also let $\tilde\lam_i$ be a relabelling of the eigenvalues $\lam_i$ of $H_1$ to include the multiplicities.
\er

In \cite[Lemma 6.5.3, p. 175]{cg:qgravity-book} we also proved
\bl\lal{5.20}
For any $\bet>0$ the operator
\begin{equation}
e^{-\bet H_1}
\end{equation}
is of trace class in $\mc H$, i.e., 
\begin{equation}
\tr (e^{-\bet H_1})=\sum_{i=0}^\un e^{-\beta \tilde\lam_i}<\un.
\end{equation}
 Let
\begin{equation}
\msc F\equiv\msc F_+(\mc H)
\end{equation}
be the symmetric Fock space generated by $\mc H$ and let
\begin{equation}
H=d\Clk(H_1)
\end{equation}
be the canonical extension of $H_1$ to $\msc F$. Then
\begin{equation}
e^{-\bet H}
\end{equation}
is also of trace class in $\msc F$
\begin{equation}\lae{5.129}
\tr(e^{-\bet H})=\prod_{i=0}^\un(1-e^{-\bet \tlam_i})^{-1}<\un,
\end{equation}
where $\tilde\lam_i$ is a relabelling of the eigenvalues $\lam_i$ to include the multiplicities.
\el
The proof relies on the fact that a temporal Hamiltonian $H_0$, which is similarly defined as the operator $\hat H_0$ in \re{5.91}, has these properties. 

For the present operator $\hat H_0$ it is also valid that $e^{-\beta \hat H_0}$ is of trace class and the proof of this property is very similar to the proof we gave in \cite[Theorem~6.2.8, p. 148]{cg:qgravity-book}, however, the structure of the operator in \re{5.91} is slightly different so that we cannot simply refer to the previous result.  We shall  give a proof in the next section instead.

\br\lar{5.21}
In \cite[Chapter 6.5]{cg:qgravity-book} we used these results to define the partition function $Z$ by
\begin{equation}
Z=\tr(e^{-\bet H})=\prod_{i=0}^\un (1-e^{-\bet\tlam_i})^{-1}
\end{equation}
and the density operator $\rho$ in $\msc F$ by
\begin{equation}
\rho=Z^{-1}e^{-\bet H}
\end{equation}
such that
\begin{equation}
\tr \rho=1.
\end{equation}

The von Neumann entropy $S$ is then defined by
\begin{equation}
\begin{aligned}
S&=-\tr(\rho\log \rho)\\
&=\log Z+\bet Z^{-1}\tr (He^{-\bet H})\\
&=\log Z-\bet\pde{\log Z}\bet\\
&\equiv \log Z +\bet E,
\end{aligned}
\end{equation}
where $E$ is the average energy
\begin{equation}
E=\tr (H\rho).
\end{equation}
$E$ can be expressed in the form
\begin{equation}
E=\sum_{i=0}^\un \frac{\tlam_i}{e^{\bet\tlam_i}-1}.
\end{equation}
Here, we also set the Boltzmann constant 
\begin{equation}
k_B=1.
\end{equation}
The parameter $\bet$ is supposed to be the inverse of the absolute temperature $T$
\begin{equation}
\bet=T^{-1}.
\end{equation}
For a more detailed analysis and especially for the dependence on $\Lam$ we refer  to  \cite[Chapter 6.5]{cg:qgravity-book}. 
\er

\section{Trace class estimates for $e^{-\beta \hat H_0}$}\las 6

Let us first consider the operator 
\begin{equation}\lae{6.1}
H_0u=\f_0^{-1}\big\{-(\frac\pa{\pa t}\big (t \pde ut\big )-t^{-1}\bar\mu u+t^3 m_2 \abs\Lam u\big\}\qq\A\, u\in C^\un_c(I)
\end{equation}
which is unitarily equivalent to the operator in \fre{5.91}. $H_0$  is essentially self-adjoint in
\begin{equation}
\mc H= L^2(\R[]_+,d\mu),
\end{equation}
where
\begin{equation} 
d\mu=\f_0dt
\end{equation}
with
\begin{equation}
\f_0(t)=t^{3-\frac4n}.
\end{equation}
We shall use the same symbol for its closure, i.e., we shall assume that $H_0$ is self-adjoint in $\mc H$ with eigenvectors $u_i\in \mc H_2$, \cf  the remarks following \fre{5.70},  and with eigenvalues $\lam_i$ satisfying the statements in \frt{5.5}, where now we denote the eigenvectors by $u_i$, since they will not be transformed.
\br\lar{6.1}
The norm
\begin{equation}
\spd {H_0u}u^\frac12
\end{equation}
is equivalent to the norm $\norm u_2$ in $\mc H_2$, \cf  \re{5.12.1} and \fre{5.13}.

Let us also assume that all Hilbert spaces are complex vector spaces with a positive definite sesquilinear form (hermitian scalar product).
\er

We shall now prove that 
\begin{equation}
e^{-\bet H_0},\qq\bet>0,
\end{equation}
is of trace class in $\mc H$. The proof is essentially the proof given in \cite[Chapter 6.2]{cg:qgravity-book} with the necessary modifications due to the different structure of the operator. 

First, we need two lemmata:
\bl\lal{6.2.2}
The embedding
\begin{equation}
j:\mc H_2\hra \mc H_0=L^2(\R[]_+,d\tilde\mu),
\end{equation}
where
\begin{equation}
d\tilde\mu =(1+t)^{-2}dt, 
\end{equation}
is Hilbert-Schmidt, i.e., for any ONB $(e_i)$ in $\mc H_2$ the sum
\begin{equation}\lae{6.9}
\sum_{i=0}^\un\norm{j(e_i)}_0^2<\un
\end{equation}
is finite, where $\norm{\cdot}_0$ is the norm in $\mc H_0$. The square root of the left-hand side of \re{6.9} is known as the Hilbert-Schmidt norm  $\abs j$ of $j$ and  it is independent of the ONB.   
\el

\bp
Let $w\in \mc H_2$, then, assuming $w$ is real valued, 
\begin{equation}\lae{6.2.23}
\begin{aligned}
\abs{w(t)}^2&=2\int_0^t\dot ww\le \int_o^\un t\abs{\dot w}^2+\int_0^\un t^{-1}\abs{w}^2\\
&\le c \norm w_2^2
\end{aligned}
\end{equation}
for all $t>0$, where $\norm\cdot_2$ is the norm in $\mc  H_2$. To derive the last inequality in \re{6.2.23} we used \re{5.12.1} and \fre{5.1}. The estimate 
\begin{equation}
\abs{w(t)}\le c\norm w_2\qq\A\, t>0
\end{equation}
is of course also valid for complex valued functions from which infer that, for any $t>0$, the linear form
\begin{equation}
w\ra w(t), \qq w\in\mc H_2,
\end{equation}
is continuous, hence it can be expressed as
\begin{equation}
w(t)=\spd{\f_t}w,
\end{equation}
where
\begin{equation}
\f_t\in\mc H_2
\end{equation}
and
\begin{equation}
\norm{\f_t}_2\le c.
\end{equation}
Now, let
\begin{equation}
e_i\in \mc H_2
\end{equation}
be an ONB, then
\begin{equation}
\begin{aligned}
\sum_{i=0}^\un\abs{e_i(t)}^2=\sum_{i=0}^\un\abs{\spd{\f_t}{e_i}}^2=\norm{\f_t}_2^2\le c^2.
\end{aligned}
\end{equation}
Integrating this inequality over $\R[]_+$ with respect to $d\tilde\mu$ we infer
\begin{equation}
\sum_{i=0}^\un\int_0^\un\abs{e_i(t)}^2d\tilde\mu\le c^2
\end{equation}
completing the proof of the lemma.
\ep
\bl\lal{6.3}
Let $u_i$ be the eigenfunctions of $H_0$, then there exist positive constants $c$ and $\ga$ such that
\begin{equation}\lae{6.19}
\norm{u_i}_2\le c\abs{1+\lam_i}^\ga\norm{u_i}_0\qq\A\, i\in\N,
\end{equation}
where $\norm\cdot_0$ is the norm in $\mc H_0$. 
\el
 \bp
 We have
 \begin{equation}
\spd{H_0u_i}{u_i}=\lam_i\spd{u_i}{u_i}
\end{equation}
 and hence,  in view of \rr{6.1},
 \begin{equation}\lae{6.21}
\begin{aligned}
\norm{u_i}_2^2&\le c_1\lam_i\int_0^\un \f_0(t)\abs{u_i}^2\\
&\le c_1\lam_i\bigg\{\int_0^{1}\f_0(t)\abs{u_i}^2+c_2\int_{1}^\un t^{3-\frac4n}\abs{u_i}^2\bigg\}. 
\end{aligned}
\end{equation} 
To estimate the second integral in the braces let us define $p=3$ and observe that
\begin{equation}
{3-\frac4n}\le p-\frac{p}n,
\end{equation}
and hence,
\begin{equation}
t^{3-\frac4n}\le t^{p-\frac p n}\qq\A\,t\ge 1.
\end{equation}
Then, choosing small positive constants $\de$ and $\e$, we apply Young's inequality, with
\begin{equation}
q=\frac p{p-p\de}=\frac 1{1-\de}
\end{equation}
and
\begin{equation}
q'=\de^{-1}
\end{equation}
to estimate the integral from above by
\begin{equation}
\begin{aligned}
\frac1q\e^q\int_{1}^\un\big\{t^{p-\frac pn}&(1+t)^{\frac pn -p\de}\big\}^q\abs{u_i}^2\\
&+\frac 1{q'}\e^{-q'}\int_{1}^\un (1+t)^{-(\frac pn -p\de)q'}\abs{u_i}^2.
\end{aligned}
\end{equation}
Choosing, now, $\de$ so small such that
\begin{equation}
(\frac pn -p\de)\de^{-1}>2
\end{equation}
the preceding integrals can be estimated from above by
\begin{equation}
\begin{aligned}
\frac 1q\e^q\int_{1}^\un (1+t)^p\abs{u_i}^2+\frac 1{q'} \e^{-q'}\int_0^\un (1+t)^{-2}\abs{u_i}^2
\end{aligned}
\end{equation}
which in turn can be estimated by
\begin{equation}
\frac 1q \e^q c \norm{u_i}_2^2+\frac 1{q'}\e^{-q'}\norm{u_i}_0^2,
\end{equation} 
in view of  \rr{6.1}.

The first integral in the braces on the right-hand side of \re{6.21} can be estimated by
\begin{equation}
\begin{aligned}
\int_0^{1}\f_0(t)\abs{u_i}^2&\le \frac 12 c\e^2\int_0^1\abs{u_i}^2\\
&\qq+\frac12\e^{-2}\int_0^\un (1+t)^{-2}\abs{u_i}^2\\
&\le \tilde c\e^2 \norm{u_i}_2^2+\frac12\e^{-2}\norm{u_i}_0^2.
\end{aligned}
\end{equation}
Choosing now $\e,\ga$ and $c$ appropriately the result follows.
 \ep
We are now ready to prove:
\bt\lat{6.4}
Let $\bet>0$, then the operator
\begin{equation}
e^{-\bet H_0}
\end{equation}
is of trace class in $\mc H$, i.e.,
\begin{equation}
\tr(e^{-\bet H_0})=\sum_{i=0}^\un e^{-\bet\lam_i}=c(\bet)<\un.
\end{equation}
\et
\bp
In view of \rl{6.2.2} the embedding 
\begin{equation}
j:\mc H_2\hra\mc H_0
\end{equation}
is Hilbert-Schmidt. Let 
\begin{equation}
u_i\in \mc H
\end{equation}
be an ONB of eigenfunctions, then
\begin{equation}
\begin{aligned}
e^{-\bet\lam_i}&=e^{-\bet\lam_i}\norm{u_i}^2\le e^{-\bet\lam_i}c\lam_i^{-1}\norm{u_i}_2^2\\
&\le e^{-\bet\lam_i}\lam_i^{-1}c\abs{\lam_i+1}^{2\ga}\norm{u_i}_0^2,
\end{aligned}
\end{equation}
in view of \re{6.19}, but
\begin{equation}
\begin{aligned}
\norm{u_i}_0^2=\norm{u_i}_2^2\,\norm{\tilde u_i}_0^2\le c\lam_i\norm{\tilde u_i}_0^2,
\end{aligned}
\end{equation}
where
\begin{equation}
\tilde u_i=u_i \norm{u_i}_2^{-1}
\end{equation}
is an ONB in $\mc H_2$, yielding
\begin{equation}
\sum_{i=0}^\un e^{-\bet\lam_i}\le  c_\bet \sum_{i=0}^\un\norm{\tilde u_i}_0^2<\un,
\end{equation}
since $j$ is Hilbert-Schmidt. Here we also used that $\lam_0>0$.
\ep

Since the operator in \fre{5.91} has the same eigenvalues as the operator in \re{6.1} we have also proved:
\bt
The operator $\hat H_0$ in \re{5.91}, which is self-adjoint in the Hilbert space $\mc {\hat H}$, has the property that
\begin{equation}
e^{-\beta \hat H_0},\qq\beta >0,
\end{equation}
is of trace class in $\mc {\hat H}$.
\et

\section{Conclusions}\las 7
We quantized the full Einstein equations and found solutions to the resulting hyperbolic equation in a fiber bundle $E$ which can be expressed as a product of spatial eigenfunctions (eigendistributions) and temporal eigenfunctions. The spatial eigenfunctions form a basis in an appropriate Hilbert space while the temporal eigenfunctions are solutions to a second order ODE in $\R[]_+$. 

The base space $\socc$ with dimension $n\ge 3$ is a Cauchy hypersurface of the quantized spacetime $N$. The solutions $u$ of the hyperbolic equation in $E$ are evaluated at $(t,x,\chi(x))$, where $\chi$ is the metric of the Cauchy hypersurface. The main assumptions for proving the existence of spatial eigenfunctions that also form a basis of a Hilbert space is that  $\socc$ is a metric product  as described in \re{3.30} and \fre{3.32}, where the compact part $M_0$ of the product might in general be hidden from observations. In case of Schwarzschild and Kerr-AdS black holes being considered in \cite{cg:qbh} and \cite{cg:qbh2} these assumptions are satisfied.

For large $n$, $n\ge 17$, and negative $\Lam$ the temporal eigenfunctions are also the eigenfunctions of a self-adjoint operator, the eigenvalues are countable and either $\Lam$ plays the role of an implicit eigenvalue, \cf \frt{5.10}, or $\Lam<0$ is considered to be a fixed cosmological constant in which case the temporal eigenfunctions are eigenfunctions of a self-adjoint operator $\hat H_0$ and a subset of the spatial eigenfunctions are eigenfunctions of a self-adjoint operator $H_1$ acting in $\socc$ such that $\hat H_0$ and $H_1$ have the same eigenvalues but with different multiplicities. The operators
\begin{equation}
e^{-\bet \hat H_0}\qq\wed\qq e^{-\bet H_1}
\end{equation}
are of trace class in their respective Hilbert spaces and also in the corresponding symmetric Fock spaces. The latter result allows to define a partition function $Z$, a density operator $\rho$, the von Neumann entropy $S$ and the average energy $E$ of the quantum system, \cf \frl{5.20} and \cite[Chapter 6.5]{cg:qgravity-book}.

\bibliographystyle{hamsplain}
\providecommand{\bysame}{\leavevmode\hbox to3em{\hrulefill}\thinspace}
\providecommand{\href}[2]{#2}



\end{document}